\documentclass[12pt,a4,epsfig]{article}
\usepackage{float}
\usepackage{epsfig,latexsym} 
\usepackage{amsthm}
 
\def\laq{\ \raise 0.4ex\hbox{$<$}\kern -0.8em\lower 0.62
ex\hbox{$\sim$}\ }
\def\gaq{\ \raise 0.4ex\hbox{$>$}\kern -0.7em\lower 0.62
ex\hbox{$\sim$}\ }
\def\half{\hbox{\magstep{-1}$\frac{1}{2}$}}

\def\NPB{{\em Nucl. Phys.} B}
\def\PLB{{\em Phys. Lett.}  B}

\def\PRD{{\em Phys. Rev.} D}

\def\mh{m_{\rm Higgs}}
\def\st{\;\widetilde{t}}
\def\mw{m_W}

\renewcommand\({\left(}
\renewcommand\){\right)}
\renewcommand\[{\left[}
\renewcommand\]{\right]}

\def\lsim{\raise 0.4ex\hbox{$<$}\kern -0.8em\lower 0.62
ex\hbox{$\sim$}}
\def\gsim{\raise 0.4ex\hbox{$>$}\kern -0.7em\lower 0.62
ex\hbox{$\sim$}}

\newcommand{\ogw}{\Omega_{\rm gw}}
\newcommand{\hogw}{h_0^2\Omega_{\rm gw}}

\newcommand{\fp}{f_{\rm peak}}

\newcommand\ee{\end{equation}}
\newcommand\be{\begin{equation}}
\newcommand\eea{\end{eqnarray}}
\newcommand\bea{\begin{eqnarray}}

\newcommand\eean{\end{eqnarray*}}
\newcommand\bean{\begin{eqnarray*}}

\newcommand{\al}{$\alpha$}

\newcommand{\bsh}{\frac{\beta}{H_*}}
\newcommand{\bshd}{$\beta /H_*$}
\begin{document}
\input epsf
\font\cmss=cmss10 \font\cmsss=cmss10 at 7pt  
%
%
 
\begin{center} 
{\Large \textbf{Gravitational Waves from Electroweak Phase Transitions}} 
\end{center} 
 
\vspace{5pt} 
\renewcommand{\thefootnote}{\fnsymbol{footnote}}
\begin{center}
Riccardo Apreda$^{a}$, Michele Maggiore$^{b}$,  Alberto Nicolis$^{c}$ and Antonio Riotto$^{d} \quad$ \footnote{email: \hspace{0.3cm} 
\texttt{${}^{a}$apreda@df.unipi.it, 
${}^{b}$michele.maggiore@physics.unige.ch,\\[1mm]
\hspace*{2cm}
${}^{c}$nicolis@cibs.sns.it, 
$ {}^{d}$antonio.riotto@pd.infn.it}} \vspace{20pt} 

$^{a}$\textit{\small{Dipartimento di Fisica, via Buonarroti 2, I-56100, Pisa and INFN, sezione di Pisa, Italy}}
  
$^{b}$\textit{\small{D\'epartement de Physique Th\'eorique, Universit\'e de Gen\`eve, 24 quai Ansermet, CH-1211 Gen\`eve 4}}

$^{c}$\textit{\small{Scuola Normale Superiore, piazza dei Cavalieri, I-56125, Pisa and INFN, sez. di Pisa,  Italy}} 
 
\textit{\small{$^d$ INFN, Sezione di Padova, via Marzolo 8, Padova I-35131,  Italy}} 
\end{center} 
 
\vspace{4pt} 
 

%
\begin{center}
July, 2001
\end{center}
\vspace{4pt} 
\begin{abstract}
Gravitational waves are generated  during   first-order  
phase transitions, either by turbolence or by bubble collisions.
If the transition takes place at temperatures of the order of the
electroweak scale, the frequency of these gravitational waves is today 
just within  the band of the planned space interferometer
LISA. We present a detailed analysis of 
the production of gravitational waves during an
electroweak phase transition in different supersymmetric models
where, contrary
to the case of the Standard Model, the transition can be first order.
We find that the stochastic background  
of gravitational waves generated by bubble nucleation
can reach a maximum value $\hogw$
of order $(10^{-10} - 10^{-11})$,
which is  within the reach of  the planned
sensitivity of LISA,
while turbolence can even produce signals  at the level
$\hogw\sim 10^{-9}$. These  values of $\hogw$ are obtained in
the regions of the parameter space which can account for the
generation of the baryon asymmetry at the electroweak scale.
\end{abstract}
\vspace{4cm}
\vfill 
\vskip2pc

\def\simlt{\stackrel{<}{{}_\sim}}
\def\simgt{\stackrel{>}{{}_\sim}}
\section*{Introduction}

Gravitational wave (GW) interferometers are presently under construction,
and have as a possible target a stochastic background of GWs of cosmological 
origin
(for  reviews, see~\cite{thorne,allen,maggiore,maggiorets}).
In particular, the space interferometer LISA is expected to fly around 
2010, with an 
extremely good sensitivity for GWs with a frequency $f$ 
between $10^{-4}$~Hz and 1~Hz~\cite{LISA}. 
It is quite fortunate that this is exactly the frequency range that  
a gravitational  wave produced at a temperature $T\sim$ 100 GeV has today, 
after redshifting, and it is therefore interesting to investigate the 
intensity of GW backgrounds produced at a possible
electroweak phase transition, and to compare it with the expected sensitivity
of LISA.

Two main production  mechanisms have been proposed for a
first order  phase transition: the first is the nucleation
and collision of true vacuum bubbles 
\cite{witten,hogan,TuWi,kosowsky,turner,envelope,kamionkowski}. 
The second is the onset of turbolence as a consequence of the injection 
of energy into the primordial plasma~\cite{kamionkowski,KMK}.

These mechanisms are only effective for a strongly first order phase
transition.
For the Standard Model, the strength of the electroweak
 transition depends of the Higgs and $W$ masses, and
can be investigate perturbatively only if $m_H\ll m_W$.
The  physical situation $m_H>m_W$ requires instead non-perturbative 
lattice computations,  which have shown~\cite{pt} that 
unfortunately there is no phase transition at all, but rather a 
smooth crossover. Therefore, in the 
Standard Model, no GWs are produced by this mechanism.

The strength of  phase transitions, however, is strongly model- and 
parameter-dependent (for review, see~\cite{RS}), 
and it is therefore meaningful to investigate
whether in extensions of the Standard Model one can have a background 
of interest for LISA. In this paper we investigate the production of
GWs in supersymmetric extensions of the SM. The strength of the phase
transition has been investigated in connection with the generation of
the baryon asymmetry~\cite{riotto}, and it is known that
in the Minimal Supersymmetric Standard Model (MSSM)  a
 strong enough phase transition requires light Higgs and stop 
eigenstates~\cite{reviewquiros}.
If the Higgs is heavier than about 115 GeV, 
stronger constraints are imposed on the space of 
supersymmetric parameters. 
If one goes beyond the MSSM, introducing an additional gauge singlet in the 
Higgs sector (the so-called next-to-minimal SSM), it is known that the
strength of the transition can be further enhanced~\cite{pietroni}.

The paper is organized as follows. In section \ref{2} 
we review the physical processes
that can lead to a stochastic background of gravitational waves 
during a phase transition; in particular,
in section \ref{2c} we describe the strategy of our computation.
Sections II and III are dedicated to the analysis of GW production,
from turbolence and from bubble collisions,
in two specific 
supersymmetric models: the 
Minimal Supersymmetric Standard Model (MSSM) and the 
Next-to-Minimal Supersymmetric 
Standard Model (NMSSM); we present a detailed analysis that extends
results already reported in~\cite{AMNR}.
Our conclusions are presented in section IV.

\section{GW production at phase transition}\label{2}
\subsection{GW background from bubble collision \label{2a}}
In first order phase transitions the Universe finds itself in a
metastable state, which is separated from the true vacuum 
by a barrier in the potential
of the order parameter, usually a scalar field $\phi$.
True vacuum bubbles are then nucleated via quantum
tunneling. In strongly first order phase transition the subsequent
bubble dynamics is relatively simple: once the bubbles are nucleated,
if they are smaller than a critical size their volume energy cannot
overcome the shrinking effect of the surface tension, and they
disappear. However, as the temperature drops below a critical
temperature $T_c$, it becomes possible to nucleate bubbles that are
larger  than this critical size. These `critical bubbles'
 start expanding, until
their wall move at a speed close to the speed of light. The energy
gained in the transition from the metastable state to the ground state
is transferred to the kinetic energy of the bubble wall.
As the bubble expands, more and more regions of space convert to the
ground state, and the wall  becomes more and more energetic. At the
same time, it also becomes thinner, and therefore the energy
density stored in the wall increases very fast. 
As long as we have a single spherical bubble, this large energy of
course cannot be converted in GWs. But when two bubbles collide spherical 
symmetry is broken and we have
the condition for the liberation of a large amount of energy into
GWs. In particular, there are two possible `combustion' modes for the
two bubbles: detonation~\cite{Ste}, that basically takes place when the
boundaries propagate faster than the speed of sound, and deflagration,
when instead they move slower. In the first case there is a large
production of GWs~\cite{hogan,witten,kamionkowski}.
The dynamics of a particular phase transition is governed by the 
effective potential for the scalar field(s) driving the transition itself.
Given this effective potential, two are the basic quantities 
which play a role in the
determination of the GW background generated during a first-order
phase transition, and in which all the relevant features of the potential are
 encoded. The parameter  $\alpha$ 
gives a measure of the jump in the energy density 
experienced by the order parameter $\phi$
during the   
transition from the false to the true vacuum and  it is
the ratio between the false vacuum energy density 
and the energy density of the radiation at the transition temperature $T_*$.
 The parameter 
$\beta$ characterizes the bubble nucleation rate per unit volume, which 
can be expressed as $\Gamma=\Gamma_0 \exp(\beta t)$ \cite{turner}. 
Thus $\beta ^{-1}=\Gamma/\dot{\Gamma}$ represents roughly 
the duration of the phase transition and   the characteristic frequency of 
the 
gravitational radiation at time of production is expected to be $2\pi f \simeq \beta$. 
The stronger is the transition, the larger is $\alpha$ and the smaller
is $\beta$ because of the
larger amount of supercooling experienced by the system before the
transition. \\
Once one has determined $\alpha$ and $\beta$,
the gravitational radiation resulting from bubbles collisions 
can be calculated, for a strongly first order transition and for detonation 
combustion mode, in function of these two parameter only.
We refer the reader to  Ref.   
\cite{kamionkowski}
 for a detailed derivation and we summarize here  
only the  strategy. The stress-energy tensor for a bubble expanding as a detonation front at velocity $v$ and of 
size $R\sim \beta^{-1} v$ takes a simple form, its relevant part coming from the wall's contribution only. Its amplitude is 
proportional only to the kinetic energy density of the wall, which is a fraction $\kappa$ of the energy gained in the 
transition, namely the false vacuum energy density $\epsilon$ times the bubble's volume $R^3 \sim (\beta^{-1} v)^3$. Then 
the energy radiated during the collision of two bubbles is calculated  
from their stress-energy tensors using the envelope approximations, 
while the theory 
of relativistic combustion supported by numerical calculations gives $\kappa$
 and $v$ as a function of $\alpha$.
Taking in account the redshift from time of production up to now, and
evaluating numerically the contribution from many randomly nucleated
bubbles in a sample volume, one finally obtains the energy density of
the radiation produced. It is convenient to express it in 
terms of the dimensionless quantity 
\be
h_0^2\ogw (f)=\frac{h_0^2}{\rho_c}\,\frac{d\rho_{\rm gw}}{d\log f}\, ,
\ee
where $\rho_{\rm gw}$ is the energy density associated to GWs, 
$f$ is their frequency and $\rho_c$ is the present value of the
critical energy density, $\rho_c =3H_0^2/(8\pi G_N)$, with $H_0=100 h_0 
 $ Km/sec Mpc; $h_0$ parametrizes the uncertainty in $H_0$
(note that the combination $\hogw$ is independent of $h_0$).
LISA is expected to reach a sensitivity of order
\be
\hogw \simeq  10^{-12}\hspace{15mm} {\rm at}\,  f=1\, {\rm mHz}\, . 
\ee
At this frequency a cosmological signal could be masked by  an astrophysical
background due to unresolved compact white dwarf binaries. Its
strength is uncertain, since it depends on the rate of white dwarf
mergers and it is estimated to be~\cite{LISA}
\be
\hogw \simeq  10^{-11}\, .
\ee
At a frequency
$f\simeq 10$~mHz the LISA sensitivity is expected to be of order
$\hogw \simeq  10^{-11}$, and the astrophysical background is
expected to be below this value.

At the peak frequency $f_{\rm peak}$,
the intensity $\hogw$ of the radiation produced in bubble collisions is
given, in terms of the parameters $\alpha$ and $\beta$ discussed above, 
by~\cite{kamionkowski}
\be
\hogw  \simeq   10^{-6} \(\frac{0.7 \: \alpha + 
0.2 \sqrt{\alpha}}{1+0.7 \: \alpha} 
\cdot \frac{\alpha}{1+\alpha}\)^2  
 \(\frac{H_*}{\beta}\)^2
\(\frac{v^3}{0.25+v^3}\)\(\frac{100}{g_*}\)^{1/3}\, ,           \label{omega}
\ee
where $H_*$ and $g_*$ are respectively the Hubble parameter and the number of relativistic degrees of freedom at the time of transition and $v$ is the velocity at which the bubble 
expands. In the following we will use the value of the velocity $v=v(\alpha)$
as given in Ref. \cite{kamionkowski} for bubble detonation.

The detailed computations of $\alpha$ and $\beta$ will be the main
issue of this paper. However,
to have a first idea of the numbers involved, we note that early
analytical estimates by
Hogan~\cite{Hogan83,hogan} indicated that it is difficult to obtain
values of $H_*/\beta$ larger that ${\cal{O}}(10^{-2})$ (this estimate will be
confirmed by our numerical results)
and therefore, even for
large values of $\alpha$, one should expect a signal at most of order 
$\hogw\sim 10^{-10}$. This maximum value can be obtained only for strongly
first order phase transition. The weaker the transition, the smaller
are both $H_*/\beta$ and $\alpha$, and therefore we have further
suppressions. It is clear that, since the value $\hogw\sim 10^{-10}$
is already close to the limiting sensitivity of LISA, one can hope to
find an observable signal only in some specific models, and in
special regions of the parameter space; an observable signal will
not be a generic property of extensions of the
Standard Model. However, we know that the condition for strongly first
order phase transition is the same that is required for producing the
baryon asymmetry at the weak scale. Therefore the models and the
regions of parameter space which produce
an interesting GW signal, albeit quite specific, are also the most
interesting if the baryon asymmetry has been indeed generated at the
weak scale, and it makes sense to perform a detailed scanning of the
parameter space in search of the strongest GW signal.

The present ({\it i.e.}  properly red-shifted) 
peak frequency of the GW background is determined by $\beta$, by
$H_*$, $g_*$, and by the transition temperature $T_*$
\cite{kamionkowski},
\be\label{peakf}
\fp \simeq 5.2 \times 10^{-6} \:  \(\frac{\beta}{H_*} \) \(\frac{T_*}{100\: {\rm GeV}} \) \( \frac{g_*}{100}\)^{1/6}  \: {\rm Hz}\:. \label{fpicco}
\ee 
At frequencies lower than $\fp$ the energy density increases 
like $f^{2.5}$, while above drops off more slowly~\cite{envelope}.
As it will be shown in the following, 
typical values for $\beta / H_*$ for the electroweak phase transition 
are  between $10^2$ and a few times $10^3$, with $T_* \sim 100$~GeV. This gives
a  frequency $\fp$ in the range $(10^{-4}- 5 \times 10^{-3})$ Hz, which 
is precisely the 
range in which LISA achieves its maximum sensitivity.
\subsection{GW background from turbolence \label{2b}}

During a first order transition there is another very general
and possibly powerful source of 
gravitational waves. Part of the energy
gained in the transition from the metastable state to the ground state
is used to heat up the plasma, and another fraction  is converted into
bulk motion of the fluid.  If the Reynolds number of the Universe at
the phase transition is large enough, then this results in the onset of
turbolence in the plasma, and consequently in the production of GWs.
To compute accurately the amount of gravitational waves pruduced by
turbolence is certainly a very difficult task. In the following, we 
will use the
estimates for the characteristic frequency and for the intensity 
given in ref.~\cite{kamionkowski},
\begin{eqnarray}
&& f_0 \simeq  2.6\times 10^{-6}\,{\rm Hz}\,  \frac{v_0}{v}\(\frac{\beta}{H_*}\)
\(\frac{T_*}{100 \:{\rm GeV}}\)\(\frac{g_*}{100}\)^{1/6}	\label{turbof}\\
&& \hogw  \simeq  10^{-5}\(\frac{H_*}{\beta}\)^2vv_0^6		
\(\frac{100}{g_*}\)^{1/3}	\;	 .\label{turbo}
\end{eqnarray}
Here $v_0$ is the fluid velocity on the largest length scales on
which the turbolence is being driven, and $v$ is again the velocity of the
bubble wall. 
For a detonation, $v_0$ can be estimated to be $v_0 \sim \sqrt{ \kappa \alpha}$ for a weak 
transition (small $\alpha$) and $v_0 \sim 1$ for a strong transition 
\cite{kamionkowski}.
Strictly speaking, eq. (\ref{turbo}) is derived in the regime of 
nonrelativistic fluid velocities
(small $v_0$). A more detailed analysis of the GW production by
turbolence has been recently announced as in preparation~\cite{KMK}.

Substituting again typical \bshd values of order $10^2 - 10^3$, 
and assuming large enough $v_0$ 
(detonation), we obtain a peak frequency around 
$10^{-4} - 5\times 10^{-4}$ Hz, again well within the window of sensitivity of 
LISA. This is therefore another,  very interesting, mechanism.

\subsection{Computational method \label{2c}}
Before analysing specific models, we  
 describe how one can compute the 
relevant quantities appearing in Eqs.~(\ref{omega}) and (\ref{fpicco}), 
namely $T_*$, \al\ and $\beta / H_*$.
We will perform a perturbative computation based
on the thermal effective potential. This method
 is known to receive large non-perturbative 
corrections when the phase transition is radiatively
induced~\cite{Strumia1,Strumia2,Moore1,Moore2,Strumia3}. In this case
 our results are really an upper bound on the GW production. However,
 the only situation in which we will find a signal compatible
 with LISA is when the potential has a metastable minimum already at
 tree level, and in this case our results are not expected to get
 important non-perturbative modifications.

The typical temperature dependence of  a potential $V(\phi,T)$
for the Higgs scalar field(s), for a 
first-order transition, is shown  in Fig.~\ref{andpot1loop} (the figure refers 
for simplicity to one-dimensional potential, but the following discussion
 is valid also if several scalar fields are
 involved).
\setlength{\unitlength}{0.15in} 
\begin{figure}[H]
\begin{picture}(28,18)(0,0)
\put(9,0.5){ 
\centering\leavevmode\epsfxsize=4.2in\epsfbox{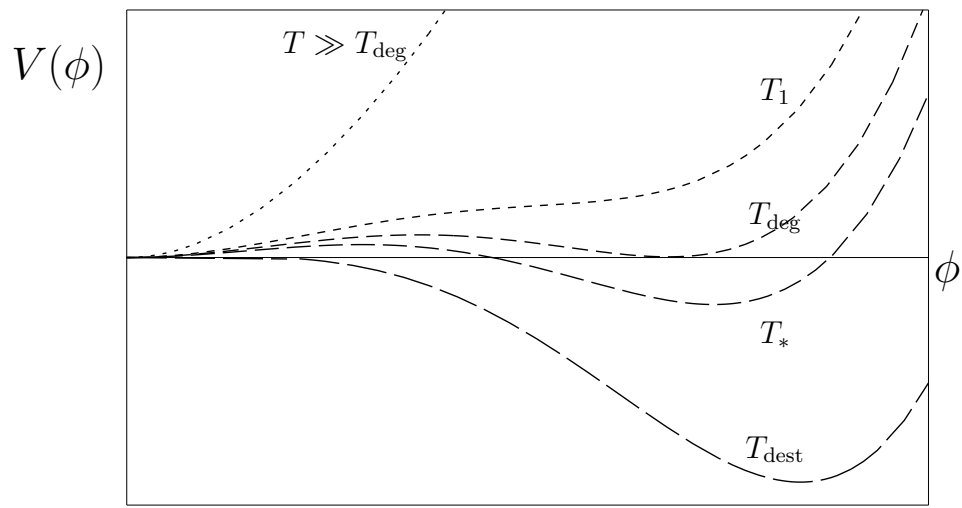}
}
\put(17,16){\makebox(0,0)[b]{$T \gg T_{\rm deg}$}}
\put(32,14.5){\makebox(0,0)[b]{$T_1$}}
\put(32,10){\makebox(0,0)[b]{$T_{\rm deg}$}}
\put(32,6){\makebox(0,0)[b]{$T_*$}}
\put(32,2){\makebox(0,0)[b]{$T_{\rm dest}$}}
\put(7,15){\makebox(0,0)[b]{\Large{$V(\phi)$}}}
\put(38,8){\makebox(0,0)[b]{\Large{$\phi$}}}
\end{picture}
\caption{Typical temperature-dependence of the potential $V$ for a scalar field $\phi$ 
driving a first order transition. The dotted curve refers to very high temperature, while
longer dashed lines refer to lower temperatures.
$T_1$ is the temperature at which a minimum at $\langle \phi \rangle \neq 0$ develops;
at $T_{\rm deg}$ the two minima are degenerate; at $T_{\rm dest}$ the origin $\langle \phi \rangle = 0$
becomes unstable. The actual transition temperature $T_*$ is between $T_{\rm dest}$ and $T_{\rm deg}$.     
\label{andpot1loop}}
\end{figure}

In the case of a first order transition 
the rate of tunneling per unit volume $\Gamma$ from the metastable
minimum to the stable one is in general suppressed by the exponential of an effective 
action, $\Gamma = \Gamma_0 \; e^{-S}\;$ with
\be             \label{azione}
S= \int d\tau d^3 x  \[ \half  \left( \frac{
 d \phi}{d \tau}\right)^2 +\half  (\vec{\nabla} \phi)^2+  V(\phi)  \]	\; ,
\ee
where $\tau$ is the euclidean time.
In the case of field theory at finite temperture $T$, one is led to consider Euclidean field theory 
periodic in imaginary time with period $T^{-1}$; moreover, in a cosmological context
the temperature $T$ of the Universe decreases in time. Thus in our case the action (\ref{azione}) 
acquires a dependence on cosmic time $t$, and has to be computed 
in the space of functions periodic in Euclidean time $\tau$, 
\be
S(t)= \int_0^{\frac{1}{T}} \! d\tau d^3 x  \[ \half  \left( \frac{
 d \phi}{d \tau}\right)^2 \! \!+\half  (\vec{\nabla} \phi)^2+  V(\phi,T)  \] \;  , \label{aztfin}
\ee
where $V(\phi, T)$ is the effective potential, shifted in such a way that 
$V(0,T) = 0$.
One then expands $S(t)$ around the transition
 time $t_*$, $S(t) \simeq S(t_*) - \beta (t-t_*)$ and identifies
$\beta$ with $- (d S / d t)_{t_*}$. From $\Gamma \propto \exp (-S(t))$
it then follows that  $\Gamma \propto \exp (\beta t)$.
In a radiation dominated Universe this gives
\be
\bsh = \left. T_*\;\frac{d \( S_3/T \)}{d T} \right|_{T_*}	\; ,	\label{defbsuh}
\ee
where we have considered the large $T$ limit  of (\ref{aztfin}) and have defined
$S_3$ as the spatial Euclidean action
\be
S_3(T)=\int d^3 x \[ \half  (\vec{\nabla} \phi_b)^2 + V(\phi_b,T)  \]
\ee
computed for the configuration of the scalar field(s) $\phi_b$ describing the
bubble.
The rate per unit volume of nucleation of a critical bubble is
therefore 
\be 
\Gamma = \Gamma_0 \; \exp \left\{ - \frac{S_3(T)}{T} \right\}\;  \label{gamma},
\ee 
where $\Gamma_0$ is of the order of $T^4$ for a finite temperature field theory.

\begin{figure}
\centering\leavevmode\epsfxsize=4.5in\epsfbox{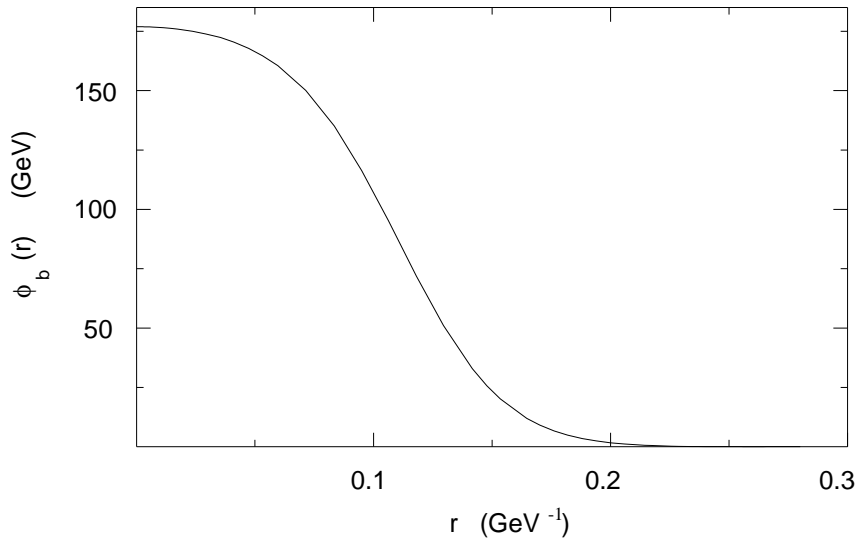}
\caption{Typical profile of a bubble solution $\phi_b (r)$ interpolating from the
true vacuum phase at $r=0$ to the false vacuum at $r=\infty$.
The solution shown corresponds to the critical bubble at the transition temperature 
in the MSSM for $\sin^2 \beta_{{\rm MSSM}} = 0.8$, $\mh=110$ GeV and $m_{{\rm stop}}=140$ GeV, see section II. 
\label{profilo}}
\end{figure}

\begin{figure}
$\qquad \; $
\centering\leavevmode\epsfxsize=5.2in\epsfbox{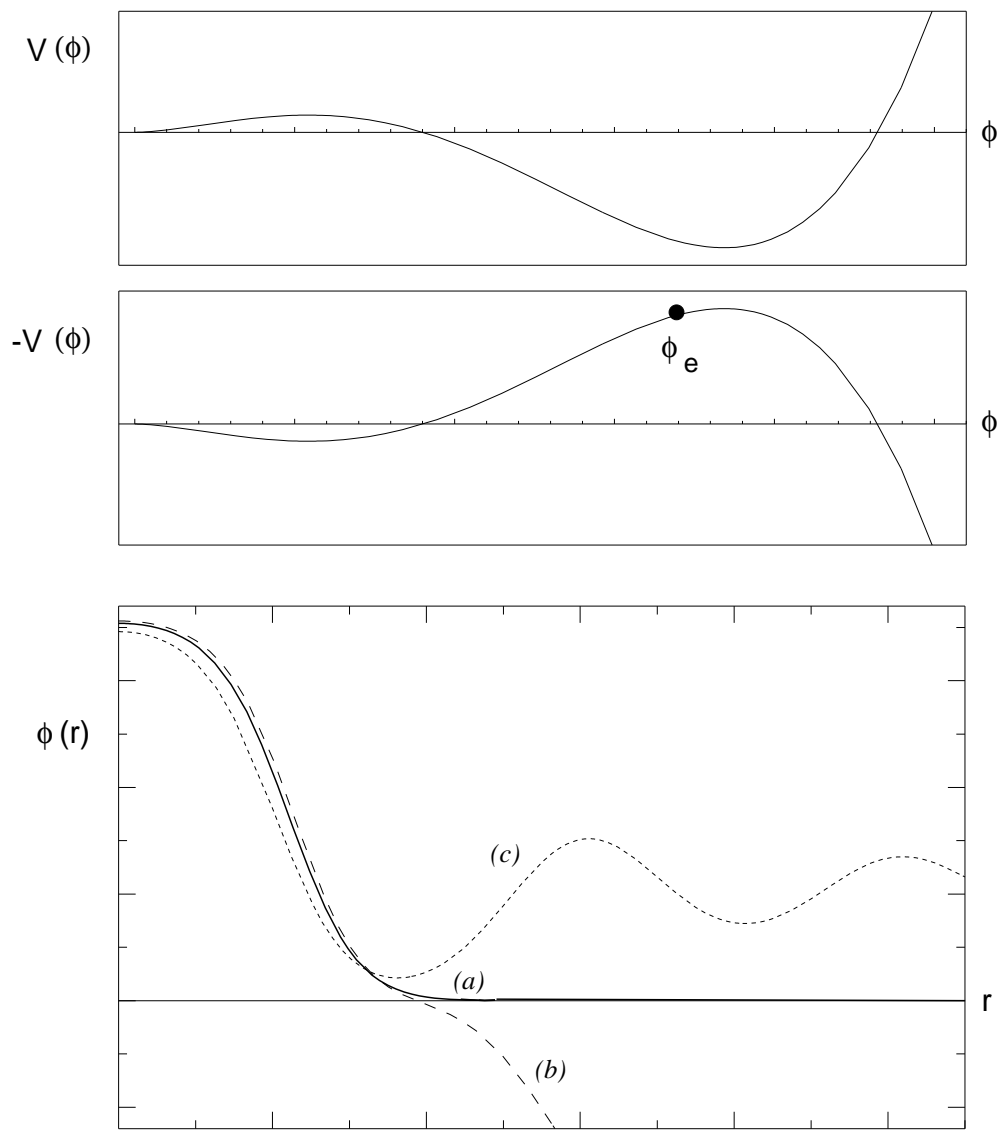}
\caption{Overshooting-undershooting method.  
\label{oushoot}}
\end{figure}

\begin{figure}
\centering\leavevmode\epsfxsize=4.8in\epsfbox{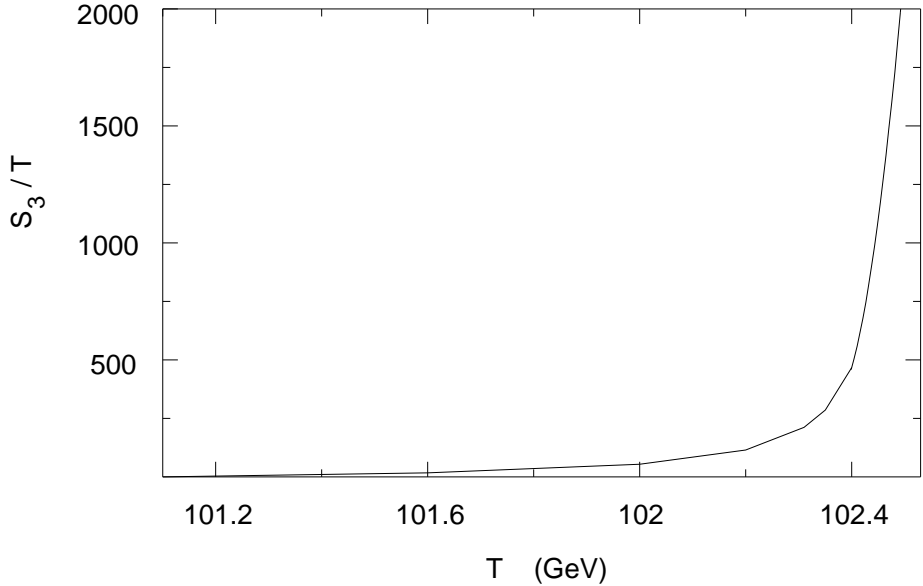}
\caption{The plot shows the rapid variation experienced by the action $S_3(T)/T$ (evaluated on critical bubble
solutions) in a small range of temperature: $S_3(T)/T$ 
moves from 0 to $\infty$ going from
$T = T_{\rm dest}$ to $T = T_{\rm deg}$,
 with $(T_{\rm deg} - T_{\rm dest}) / T_{\rm dest}
\sim 1 \%$.
The particular case shown in the figure refers to the electroweak transtion in the
MSSM with $\mh=110$ GeV, $m_{{\rm stop}}=145$ GeV, $\sin^2 \beta_{{\rm MSSM}}=0.8$, see section II. 
\label{s3sut}}
\end{figure}

The Euclidean action for a spherical configuration is
\be
S_3(T)  =
4 \pi \int dr\; r^2
\left[ \half \left( \frac{d\, \phi_b}{d\, r} \right )^2  +
       V(\phi_b, T)
\right]				\; ;
\label{accion}
\ee
the bubble configuration is thus the solution of the equation(s)
\be
\frac{d^2 \phi_b} {d r^2} + \frac{2}{r} \frac{d \phi_b}{d r} 
 - \frac{\partial V}{\partial \phi_b} = 0
\label{eqbolla}
\ee
supplemented by the boundary conditions 
\be
\left.\frac{d \phi_b}{d r} \right|_{r=0} = 0 	\: ,  \qquad 
\left.\phi_b\right|_{r=\infty} = 0 \; ,
\label{cbbolla}
\ee
namely true vacuum inside the bubble and metastable vacuum outside.
A typical bubble profile solution is shown in Fig.~\ref{profilo}.
In order to find the ``escape point'' ({\em i.e.} the value $\phi (r=0) \equiv \phi_{\rm e}$, 
implicitily determined by the two conditions (\ref{cbbolla})) one uses the 
``overshooting-undershooting'' method illustrated in fig. \ref{oushoot}.
Equation (\ref{eqbolla}) is the classical equation of
motion for a point particle subject to a potential $-V(\phi)$ and to 
a velocity- and time-dependent
friction force, if $\phi(r)$ stands for the trajectory of the particle.
If the particle starts at rest precisely from the 
escape point, it will have just enough 
energy to overcome the friction force and to come at rest at $\phi=0$ at 
infinite $r$ (case ({\em a})).
If instead it starts at the right of $\phi_{\rm e}$ (overshooting),
it will continue toward $\phi=-\infty$ (case ({\em b})).  
If it starts at the left of $\phi_{\rm e}$ (case ({\em c}), undershooting) it
will experience damped 
oscillation around the minimum of the inverted potential.
Thus one determines the escape point by trials and errors, 
lowering the starting value if 
one gets a solution of type ({\em b}), and increasing it if one gets a 
solution of type ({\em c}). 
Once the escape point is found, the extremal action (\ref{accion}) 
can be computed 
for the corresponding solution.
The bubble solution only exists in the range $T_{\rm dest}<T<T_{\rm deg}$, see fig.~\ref{andpot1loop};
for a one-dimensional potential the quantity $S_3(T)/T$ behaves as shown 
in fig.~\ref{s3sut}: at $T>T_{\rm deg}$ the 
transition cannot take place, because it is not
energetically favorable to go from one minimum to the other, 
thus $S_3 \to \infty$ and 
$\Gamma \to 0$. Conversely,  for $T$ approaching $T_{\rm dest}$ the 
transition becomes more and 
more convenient, $S_3 \to 0$ and $\Gamma \to \Gamma_0$.
The temperature $T_*$ at which the
transition takes place is  computed by comparing the
probability of bubble nucleation per unit time and unit
volume (\ref{gamma}) with the expansion rate of the
universe at the corresponding temperature.
The condition that must be satisfied is that the probability for a single
bubble to be nucleated within one horizon volume is ${\cal{O}}(1)$,
\be	\label{integral}
 \int^{t_*}_0 \frac{\Gamma}{H^3} dt = 
 \int _{T_*} ^{\infty} \frac{dT}{T} 
\left(\frac{90}{8 \pi^3 \: g}\right)^2 
\left(\frac{M_{\rm Pl}}{T}\right)^4 
e^{-S_3(T)/T} \sim 1    \; .
\ee
For the electroweak transition, in which characteristic temperatures are ${\cal{O}}
(100 \: {\rm GeV})$, the condition above is well approximated by
 $S_3(T_*)/T_* \simeq 140$.

Once one has found $T_*$, the parameter  \al \ is readily computed from the 
definition given above. Calling $v(T)$ the vacuum expectatuion value of the Higgs field 
at the true vacuum at temperature $T$, then the quantity  
\be
\epsilon_* = \left. - V(v(T),T) + T \frac{d}{dT} V(v(T),T) \right| _{T_*}
\ee
is the vacuum energy (latent heat) density associated with the transition (remember that the effective potential
$V(\phi, T)$ is the free energy density). Thus \al~is the ratio between 
this and the energy density of the radiation,
\be     \label{defalpha}
\alpha= \frac{30 \epsilon_*}{\pi^2 g_* T_*^4}   \; ,
\ee
while \bshd\  
is given  by eq.~(\ref{defbsuh}).
Notice that in order to compute \al\ it is sufficient to know the transition temperature $T_*$ 
and the potential $V(\phi, T)$, while the determination of \bshd\ is possible only if one knows
how the action $S_3(T)$ computed on the exact bubble solutions depends on $T$ in a neighbourhood of $T_*$.

\section{The Minimal SuperSymmetric Standard Model \label{sez4}}

As we have already noticed, extensions of the SM are required to 
obtain a first-order phase transition at the electroweak scale, 
because in the SM, in the physical case $\mh > \mw$, there is no phase
transition. 

The Higgs sector of the MSSM requires two complex Higgs doublets, with
opposite hypercharges
\begin{equation}
\label{higgsmssm}
H_1  =  \left(
\begin{array}{c}
H_1^0 \\
H_1^- 
\end{array}
\right)_{-1/2} \; , \qquad
H_2  =  \left(
\begin{array}{c}
H_2^+ \\
H_2^0
\end{array}
\right)_{1/2}	\; .
\end{equation}
The tree-level potential is
\be
\label{potmssm}
V^{(0)} =  m_1^2 H_1^\dagger H_1+m_2^2 H_2^\dagger H_2
+m_3^2(H_1 H_2+ {\rm h.c.}) 
+\frac{1}{8}g^2\left(H_2^\dagger \vec{\sigma} H_2+
H_1^\dagger\vec{\sigma}H_1\right)^2 
+\frac{1}{8}g'^2\left(H_2^\dagger H_2-H_1^\dagger H_1\right)^2
\ee
where the products between doublets are the usual SU(2) invariant products, the $\sigma_i$ are the Pauli matrices, 
$g$ and $g'$ are the gauge couplings and the $m_i^2$ are real parameters, not necessarily positive. In fact, in
order to have the gauge group SU(2)$\times$U(1) broken at zero temperature, one needs the quadratic term to be negative
along some directions in the Higgs fields space.

We define $v_1 = \langle H_1 ^0 \rangle$, $v_2 = \langle H_2 ^0 \rangle$ and 
$\tan \beta_{\rm MSSM} = v_2 / v_1$. 
In the following we will indicate  the angle $\beta_{\rm MSSM}$
simply by $\beta$: it should not be confused with
the parameter $\beta$ discussed in the previous sections that sets 
the time scale of phase transition.\\
As usual, the $W$ mass fixes $v^2 \equiv v_1^2 + v_2^2$ 
to the value $(246 \; {\rm GeV})^2$.
If the mass $m_A$ of the CP-odd
field in the Higgs sector  is much  larger than $m_W$, 
only one light Higgs scalar $\phi$ is left
and its tree-level potential is identical to the one in the SM, namely
$V_0(\phi) = -\frac{m^2}{2} \phi^2 + \frac{\lambda}{8} \phi^4$. This will be the case considered in
this section. When $m_A\sim m_W$
the full two-Higgs potential should be considered, but the strength of
the phase transition is weakened \cite{reviewquiros}.
As we will see in the following, the 
one-loop thermal 
corrections to the effective potential make the quadratic term
 positive at high temperature and create a negative
cubic term due to loops of the massive bosons in the theory. Due 
to the presence of this cubic term and the positivity of the
quadratic one, there exists a range of temperature in which the 
point $\phi = 0$ is a local minimum
separated from the true simmetry-breaking one by a small potential barrier,
 that is precisely 
the set-up for a first-order phase transition.  The 
strength of the  phase transition is enhanced
by the presence of  new bosons coupled to the Higgs, a significant
role being played by the right-handed  stop $\st_R$, which is 
-- apart from the Higgs itself --   the
lightest scalar  in the theory and has 
the largest Yukawa coupling to the Higgs $\phi$. 

To study the amount of gravitational waves generated during the electroweak
phase transition within the MSSM  we have made use
of the thermal potential corrected up to two-loop level \cite{carena}.
However, as discussed above, in this radiatively induced case,
non-perturbative effects are also important. The results of
ref.~\cite{Moore1} indicate that the two-loop perturbative computation
overestimates the amount of supercooling, so that the results of
this section are really  upper bounds on the GW production.

Following \cite{carena}, we summarize here
the one and two-loop finite
temperature contributions to the effective 
potential for the Higgs field $\phi$. 
At one loop,
the dominant corrections come
from the gauge bosons $Z$ and $W$, the top quark and its supersymmetric
partners. We will work in the limit in which the left handed
stop is heavy, $m_Q \simgt 500$ GeV;
lower values of $m_Q$ make the phase transition
stronger, but give large
contribution to the $\rho$-parameter.
For right-handed stop masses below (or of order
of) the top quark mass, and for large values of the CP-odd 
Higgs mass, $m_A \gg m_Z$,
the one-loop 
effective potential admits the high
temperature expansion \cite{carena}
\be
V_0(\phi) +V_1(\phi,T) = -\frac{m^2(T)}{2} \phi^2 + \frac{\lambda(T)}{8} \phi^4  \label{potMSSM} - T
\left[E_{\rm SM}\; \phi^3 +  (2 N_c) \frac{\left(m_{\rm stop}^{2} (\phi) +
\Pi_{\rm stop}(T)\right)^{3/2}}
{12 \pi} \right]	\; ,
\ee
where the various quantities are defined as follows:
$N_c=3$ is the number of colors; 
$E_{\rm SM}$ is the cubic term coefficient in the Standard Model case,
\be
\label{ESM}
E_{\rm SM} \simeq \frac{1}{3}\
\left(\frac{2 m_W^3+ m_Z^3}{2 \pi v^3}\right) \; , 
\ee
with $m_W$ and $m_Z$ standing for 
the physical masses of the gauge bosons at zero temperature.
Here $m_{\rm stop} (\phi)$ is the lightest stop mass, which depends 
explicitely on the VEV 
of the Higgs $\phi$ and is approximately given by
\begin{equation}        \label{mstop}
m_{\rm stop}^2 (\phi) \simeq m_U^2 + \left[ 0.15 \frac{m_Z^2}{v^2} \cos2\beta
+ \frac{m_t^2}{v^2} 
\left(1- \frac{\tilde{A}_t^2}{m_Q^2}\right) \right] \phi^2              \; ,
\end{equation}
where $m_t$ is the zero-temperature top quark mass, $m_U^2$ 
is a parameter of the model -- the 
soft mass squared, not necessarily positive -- and $\tilde{A}_t$ is the stop mixing parameter (which for simplicity will be set to zero in 
our numerical analysis). Here 
$\Pi_{\rm stop}(T)$ is the finite temperature contribution to the right-handed stop 
self-energy,
\be           
\Pi_{\rm stop}(T)=\frac{4}{9}g_s^2 T^2 +
\frac{1}{6}h_t^2 \left[1+\sin^2\beta \left(1 - \widetilde{A}_t^2/m_Q^2
\right)\right] T^2 
+\left(\frac{1}{3}-\frac{1}{18}|\cos 2\beta|\right)g'^2 T^2      \; ,\label{pistop}
\ee
where $g_s$ is the strong gauge coupling
and $h_t$ is the top Yukawa coupling; in our notations
$m_t = h_t v \sin\beta/\sqrt{2}$.
Finally $m^2(T)$ is given by 
\be
\frac{1}{2}\mh^2 + \frac{T^2}{v^2} \(\frac{1}{4} \mh^2 +\frac{5}{6} \mw^2 + \frac{5}{12} m_W^2 + m_t^2 \)
\ee
and $\lambda(T)\simeq \mh^2 / v^2$ apart from logarithmic corrections.
As was observed in Ref.~\cite{CQW}, the phase transition strength
is maximized for values of the soft breaking parameter 
$m_U^2 \simeq - \Pi_{\rm stop}(T)$, for which the coefficient of the cubic
term in the effective potential,  
\be
E \simeq E_{\rm SM}+ \frac{h_t^3 \sin^3\beta \left(1 -
\widetilde{A}_t^2/m_Q^2\right)^{3/2}}{4 \sqrt{2} \pi},
\label{totalE}
\ee
which governs the strength of
the phase transition (in this case $v(T_*)/T_* \simeq 4 E / \lambda$),
can be one order of magnitude larger than $E_{\rm SM}$~\cite{CQW}.
However, it was also noticed that such large negative values of
$m_U^2$ may induce the presence of color breaking minima  because the 
effective stop mass at finite temperature $m^2_{\rm stop} (\phi) + \Pi_{\rm stop}(T)$ 
may become negative \cite{CW,CQW}. The problem  arises for small values of
the VEV of $\phi$ and of the temperature $T$, as one can easily infer from 
Eqs. (\ref{mstop})
and (\ref{pistop}).
Thus, the check one has to perform is that for $T > T_*$ -- {\em i.e.} before
 the
phase transition takes place -- the sum $m^2_{\rm stop} (\langle \phi \rangle) 
+ \Pi_{\rm stop}(T) = m_U^2 +\Pi_{\rm stop}(T)$ is positive, the equality being
 motivated by the fact that $\langle \phi \rangle =0$ for $T > T_*$.
Once the above condition is satisfied, the color breaking minima problem is 
absent
for $T < T_*$ too, because the VEV $\langle \phi \rangle$ gets a large 
value ${\cal{O}} (100 \: {\rm GeV})$ after the phase transition.

Before performing the full computation with
the two-loop corrections to the potential, we proceed to a rough 
computation of the quantity $\alpha$ defined in section \ref{2a}, 
using the one-loop effective potential (\ref{potMSSM}), 
in order to understand the
dependence of the strength of the phase transition on the various 
parameters of the model, and to identify the regions of the parameter
space where it is appropriate to perform  more detailed computations.

\begin{figure}
\centering\leavevmode\epsfxsize=4.5in\epsfbox{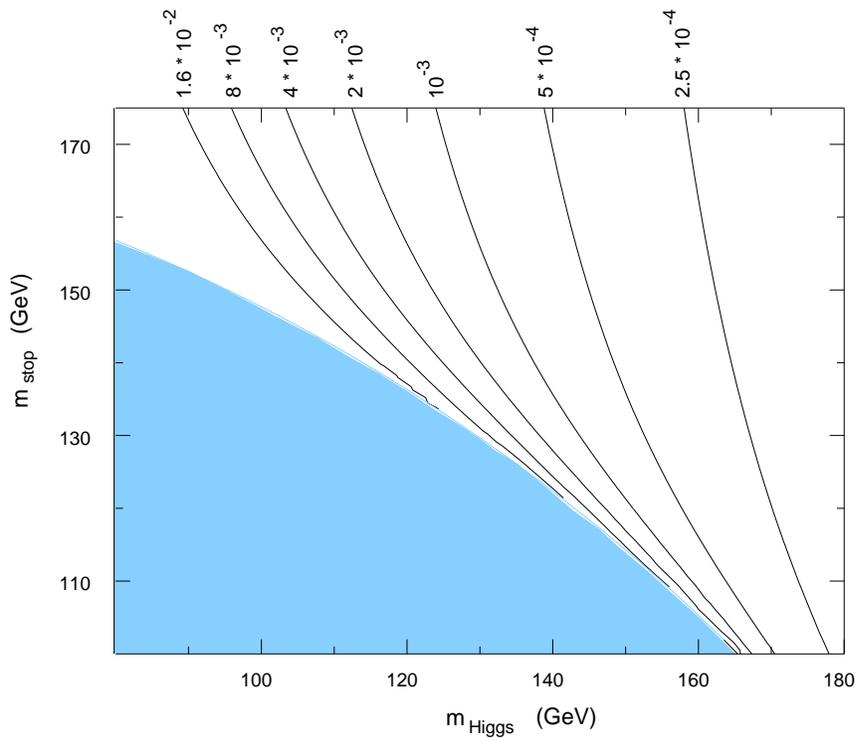}
\caption{Contour plot of $\alpha$ as a function of the Higgs and stop masses 
for $\sin^2 \beta = 0.8$. The shaded region is forbidden by the
request of absence of color breaking minima. $\alpha$ is computed at the
destabilization temperature, as discussed in the text. \ \ \ \ \ \ \ \ \ \ \ 
\ \ \ \ \ \ \ \ \ \ \  ${}^{}$ 
 \label{MSSMcontour}}
\end{figure}

\begin{figure}
\centering\leavevmode\epsfxsize=4.5in\epsfbox{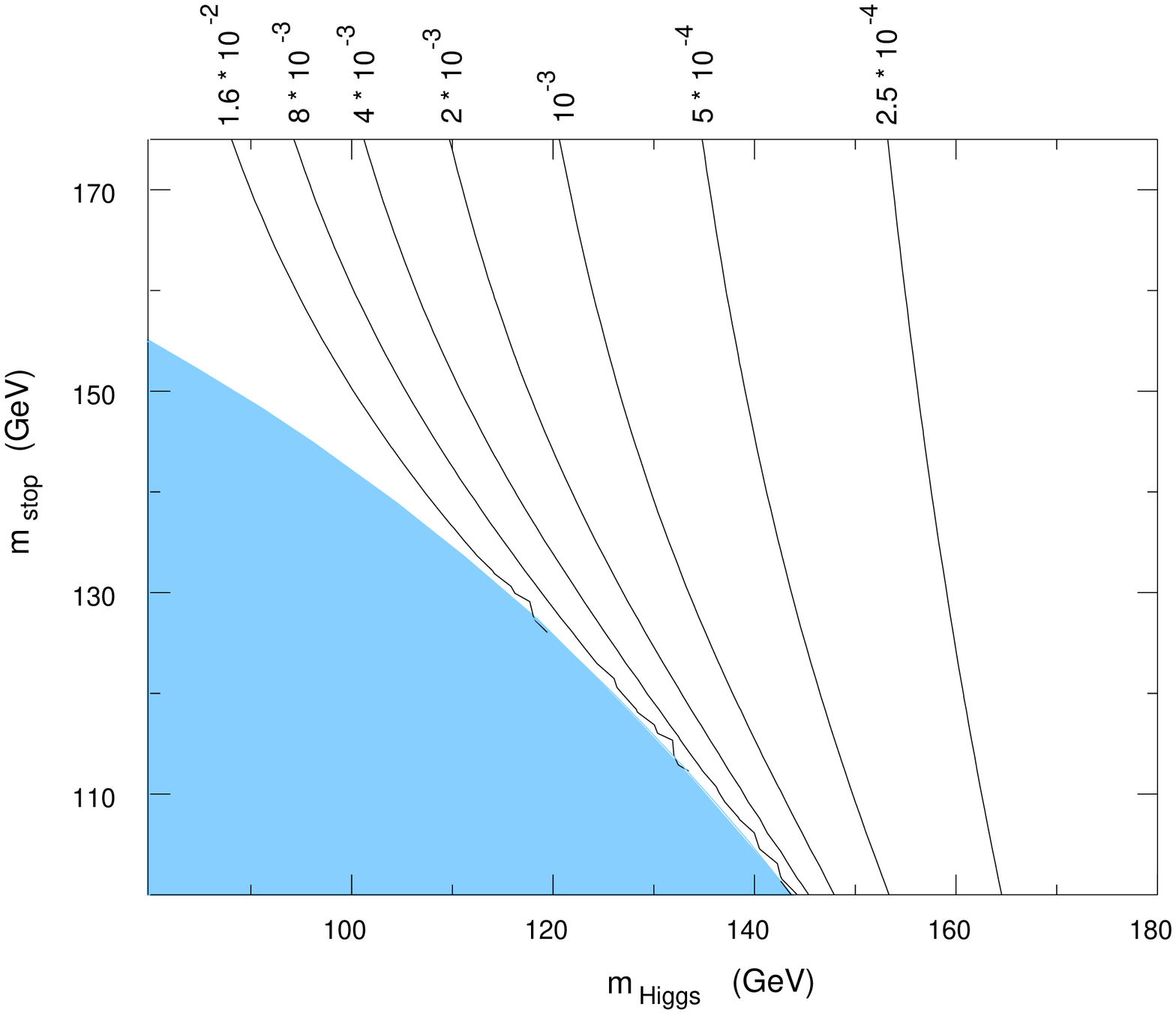}
\caption{Contour plot of $\alpha$ as a function of the Higgs and stop masses for $\sin^2 \beta = 0.3$
. \label{MSSMcontour2}}
\end{figure}

Under the assumptions mentioned above, the most relevant parameters in the game  are the Higgs mass 
$m_{\rm Higgs}$, the right-handed stop mass $m_{\rm stop}$
and the zero temperature ratio between the vacuum expectation values
of the two neutral Higgses $\tan\beta$. However, it is well known 
that the
strength of the transition has a very slight dependence on $\tan\beta$, with slightly stronger transition at high 
({\em i.e.} greater than about 0.6) values of $\sin^2 \beta$.
In our computations we will set both $\sin ^2 \beta = 0.8$ and $\sin ^2 \beta = 0.3$, obtaining very similar results.
Our strategy has been the following. For every value of $m_{\rm Higgs}$ and 
$m_{\rm stop}$ we
computed the temperature $T_{\rm dest}$ at which the origin $\phi=0$ gets destabilized;
 then
we computed $\alpha$ setting $T_* = T_{\rm dest}$ in the definition (\ref{defalpha}). 
The actual
temperature $T_*$ is in the range between the ``destabilization temperature'' 
$T_{\rm dest}$ and the
``degeneracy temperature'' $T_{\rm deg}$, as discussed in section \ref{2c}. For 
the model
under study this range is very small, $(T_{\rm deg} - T_{\rm dest}) / T_{\rm dest} \sim (1 - 3) \, \%$.
So even taking $T_*$ to be equal to $T_{\rm dest}$ leads nevertheless to the right order of magnitude in the calculation 
of the parameter \al. 
Our results are summarized in figs.~\ref{MSSMcontour} and \ref{MSSMcontour2}, which show contour 
plots for
$\alpha$ as a function of $m_{\rm Higgs}$ and $m_{\rm stop}$, together with 
the region forbidden by the condition of absence of color breaking minima 
discussed above.
From figs.~\ref{MSSMcontour} and \ref{MSSMcontour2} we see that in order to have large values for 
$\alpha$ one needs small Higgs and stop masses; this agrees with the well known fact that  the strength of the 
electroweak transition is enhanced by taking light Higgs and stop. We see also that the strength of the transition is
very slightly dependent on $\sin ^2 \beta$; thus for definiteness we fix $\sin ^2 \beta= 0.8$.

The most important two loop corrections are of the form
$\phi^2 \log(\phi)$ and are induced by 
the Standard Model weak gauge bosons as well as by the stop and
gluon loops \cite{twoloop,JoseR}
\be
V_2(\phi,T) \simeq  \frac{\phi^2 T^2}{32 \pi^2}
\left[\frac{51}{16}g^2  - 3 h_t^4 \sin^4\beta \left(1- \frac{\widetilde{A}_t^2}{m_Q^2}\right)^2
+  8 g_s^2 h_t^2 \sin^2\beta \left(1- \frac{\widetilde{A}_t^2}{m_Q^2}\right)
\right] \log\left(\frac{\Lambda_H}{\phi}\right) 
\ee
where the first term comes from the Standard Model
gauge boson-loop contributions, 
while the second and third terms come from the
light supersymmetric particle loop contributions.
The scale 
$\Lambda_H$ depends on the finite corrections \cite{twoloop} and is of 
order of 100 GeV: given the slight logarithmic dependence of $V_2$ on $\Lambda_H$, in the 
following we will set $\Lambda_H = 100$ GeV.
The complete potential we use is thus $V(\phi,T) = V_0 (\phi) + V_1(\phi,T)
+ V_2(\phi,T)$.

\begin{figure}[H]
\centering\leavevmode\epsfxsize=4.5in\epsfbox{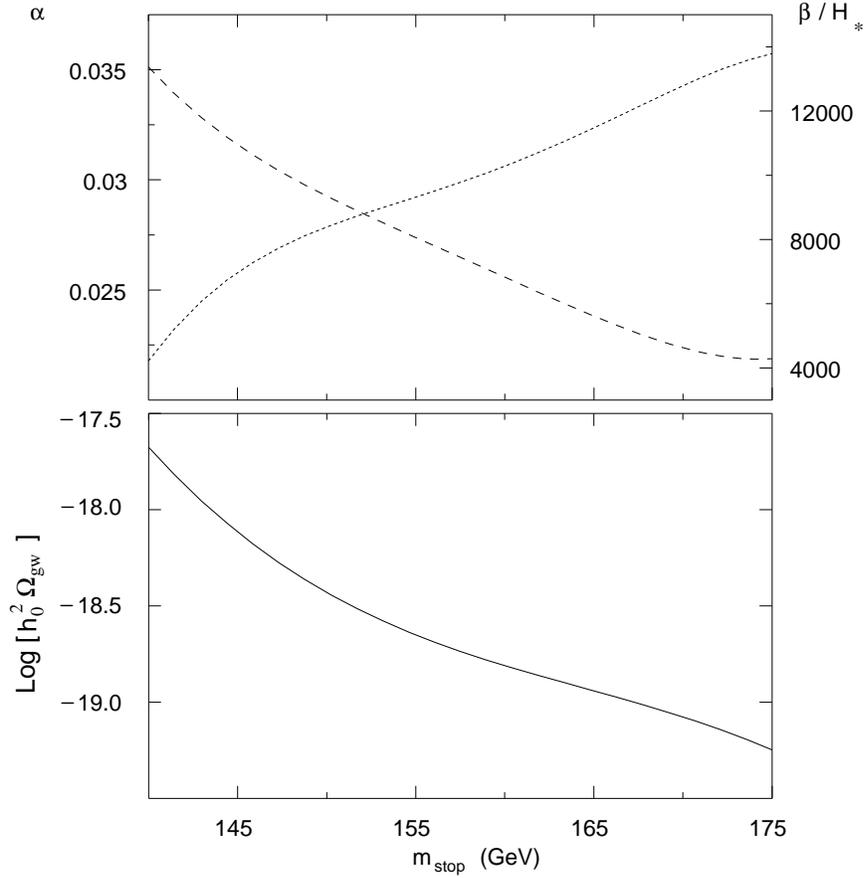}
\caption{Plot for \al\ (dashed line), $\beta / H_*$ (dotted line),
 and $\hogw$ from bubble collisions  (solid line), 
as functions of the stop mass, 
for a 110 GeV Higgs and  $\sin^2 \beta = 0.8$.  
\label{MSSMmh}}
\end{figure}

\begin{figure}[H]
\centering\leavevmode\epsfxsize=4.5in\epsfbox{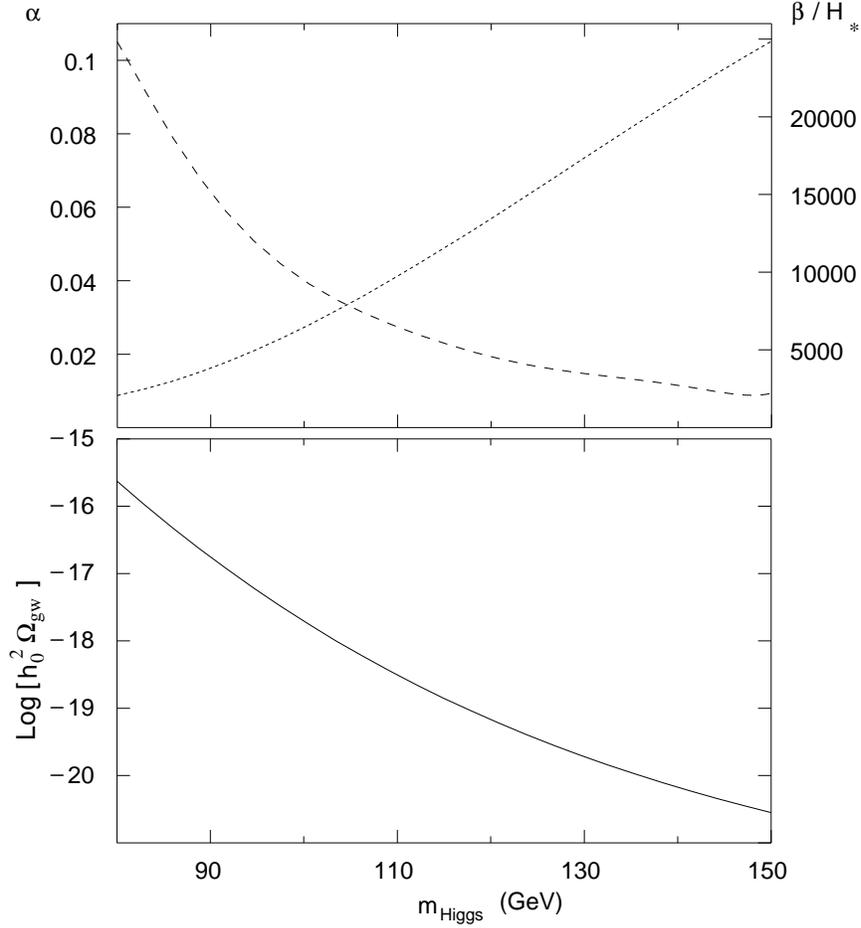}
\caption{Plot for \al\ (dashed line), $\beta / H_*$ (dotted line), and 
$\hogw$ from bubble collisions (solid line), as functions of the Higgs mass, 
for a 155 GeV stop and  $\sin^2 \beta = 0.8$. 
\label{MSSMms}}
\end{figure}

Next, having fig.~\ref{MSSMcontour} in mind, 
we proceeded to an accurate numerical computation
of the parameters $\alpha$ and $\beta / H_*$ characterizing the strength 
of the phase transition: 
for any given choice of the masses $m_{\rm Higgs}$ and $m_{\rm stop}$, we have 
first numerically computed the nucleation temperature $T_*$ by imposing that, for the
Higgs field configuration describing the nucleated bubble, 
the condition $S_3(T_*)/ T_* \simeq 140$ is satisfied. Then we have computed the parameters 
$\alpha$ and  $\beta / H_*$ through  Eqs.~(\ref{defalpha}) e (\ref{defbsuh}).
Our results are summarized in Figs.~\ref{MSSMmh} and \ref{MSSMms}.
The intensity of the produced gravity waves rapidly grows as the 
masses decrease. 
This is explained by the fact that
$\hogw$ scales roughly as  $\alpha^3/\beta^2$ and a linear increase of 
$\alpha$ and decrease of $\beta$ leads to a rapid power-law growth of the
intensity.
Unfortunately, the general prediction is that the intensity of the
gravitational waves produced during the MSSM phase transition
is too small for LISA. For instance, 
taking a Higgs mass of  110 GeV, the right-handed 
stop mass of 140 GeV and $\sin^2 \beta = 0.8$ (see fig.~\ref{MSSMmh}), we find 
$\alpha \simeq 3\times 10^{-2}$ and \bshd $\simeq 4\times 10^3$, leading
 to $\hogw \simeq 2 \times
10^{-18}$. 
Notice that one is not allowed to lower too much the
right-handed  stop mass   because of the problem of color breaking minima described above.
We have also estimated  what happens if we lower
the Higgs mass down to (the already excluded value of) 80 GeV, 
setting  the right-handed stop mass at 155 GeV -- which is 
the lower value compatible with the 
absence of color breaking minima -- and $\sin^2 \beta = 0.8$, see fig.~\ref{MSSMms}. 
We obtain  
$\alpha \simeq 0.1$ and \bshd $\simeq 2\times 10^3$, giving 
$\hogw \simeq 2 \times
10^{-16}$, a signal still not relevant. The situation does
not  improve  when both Higgses are involved in the transition because
the strength of the phase transition is weaker.
An uncertainty in this estimate is due to the determination of $v$, the velocity of expansion
of the bubble discussed in section \ref{2a}. If
the phase transition is not strong enough, then
$v$ is subsonic, so that the value of $\hogw$
is further suppressed.
\pagebreak

\begin{figure}[H]
\centering\leavevmode\epsfxsize=4.5in\epsfbox{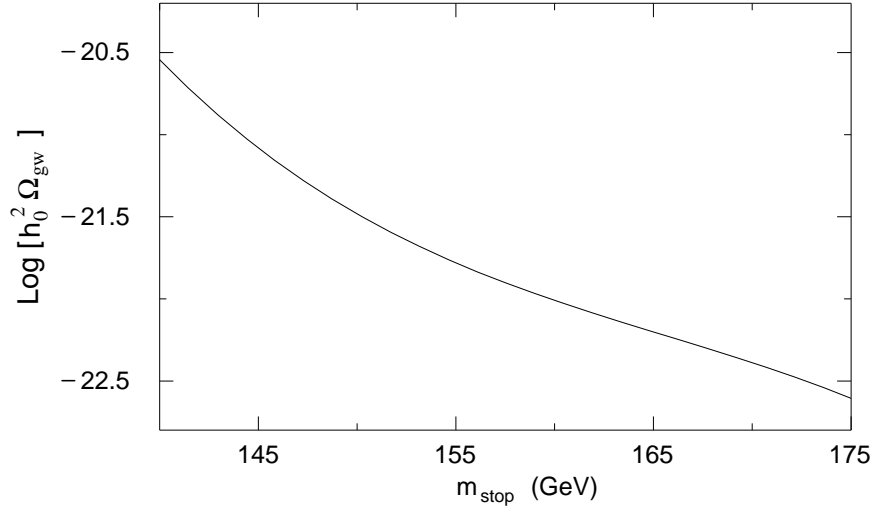}
\caption{$\hogw$ from turbolence, 
as a function of the stop mass for a 110 GeV Higgs and $\sin^2 \beta = 0.8$. \label{turbo1}}
\end{figure}

\begin{figure}[H]
\centering\leavevmode\epsfxsize=4.5in\epsfbox{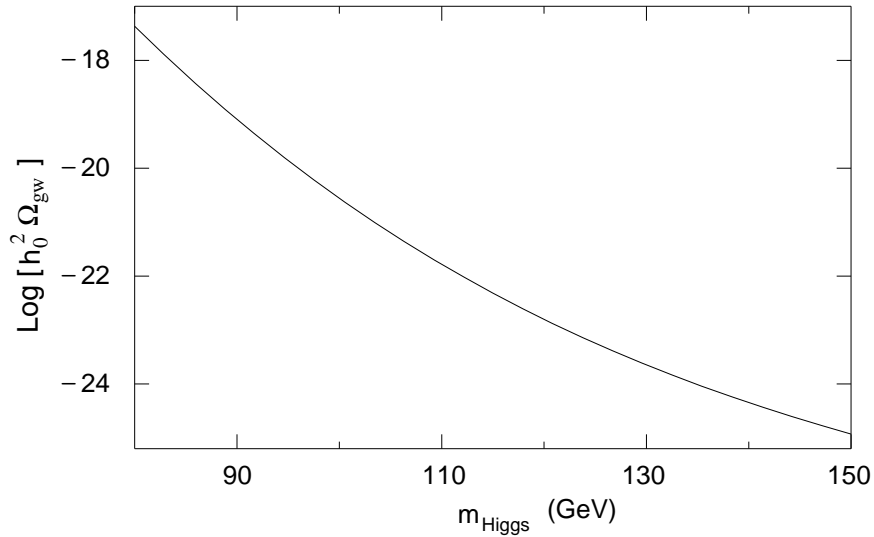}
\caption{$\hogw$ from turbolence,
as a function of the Higgs mass for a 155 GeV stop and $\sin^2 \beta = 0.8$. \label{turbo2}}
\end{figure}

It is now easy to estimate also  the amount of GWs produced by 
turbolence at the electroweak transition, in the
MSSM case. Substituting in eq. (\ref{turbo}) tha values of $\alpha$ and $\beta / H_*$ found above,
we obtain the plots shown in figs.~\ref{turbo1} and \ref{turbo2}.
Again, the results are too small to be of interest for LISA.


\section{\label{sez5}The Next-to-Minimal Supersymmetric Standard Model}

\subsection{The model}
The situation improves considerably if we enlarge the MSSM sector
adding a complex gauge singlet $N$ \cite{ellis}. This is the so-called Next-to-Minimal
Supersymmetric Standard Model (NMSSM) and 
is a 
particularly 
attractive model  to explain the observed baryon asymmetry
at the electroweak phase transition. 
The relevant part of the superpotential
is given by $$
W=\lambda H_1 H_2 N-\frac{k}{3} N^3 \;, $$
where now the supersymmetric $\mu$-parameter of the MSSM is 
substituted by the the combination
$\lambda \langle N\rangle$, and $k$ is a free parameter. 
The corresponding tree level Higgs potential reads
$V=V_{F}+V_{D}+V_{{\rm soft}}$, where 
\bea \label{pot}
V_{F} & = & |\lambda|^{2}\left[|N|^{2}(|H_{1}|^{2}+|H_{2}|^{2})+|H_{1}H_{2}|^{2} \right]
                + k^{2} |N|^{4} -(\lambda k^{*} H_{1}H_{2}{N^{2}}^{*} + 
{\rm h.c.})		 \\
V_{D} & = & \frac{g^{2}+g'^{2}}{8} (|H_{2}|^{2}-|H_{1}|^{2})^{2} +
\frac{g^{2}}{2}|H_{1}^{\dag}H_{2}|^{2}		 \\ 
V_{{\rm soft}} & = & m_{H_{1}}^{2}|H_{1}|^{2} + m_{H_{2}}^{2}|H_{2}|^{2} + m_{N}^{2}
|N|^{2} -  \left(\lambda A_{\lambda}H_{1}H_{2}N
-\frac {1}{3} k A_{k} N^{3} + {\rm h.c.}\right)		\; . 
\eea
The presence of the cubic supersymmetry breaking soft terms 
proportional  to the parameters  $A_\lambda$ and $A_k$
 already at zero temperature makes it clear that
within the NMSSM it is quite easy to get a very strong first-order
phase transition at the electroweak scale \cite{pietroni}.
The order of the transition is determined by these trilinear soft 
       terms rather than by the cubic term appearing in the
finite temperature one-loop corrections and the
 preservation of baryon
       asymmetry after the phase transition is
possible  for masses of the lightest scalar up to about 170 GeV. 
At the same time, since the transition is induced by the
tree-level terms rather than by radiative corrections,
non-perturbative effects are non expected to change
dramatically the results of the perturbative computation.

By a redefinition of the global phases of $H_{1}$ and $N$ it is always possible
 to take
$\lambda A_{\lambda}$ and $k A_{k}$ real and positive, while by an
SU(2)$\times$U(1) global rotation one can put
$v^{\pm} \equiv \langle H^{\pm}\rangle =0$ and $v_{2} \equiv \langle H_{2}^{0}\rangle \in R^{+}$.
Assuming CP conservation, also
$v_1 ^0 \equiv \langle H_{1}\rangle$ and $x \equiv \langle N\rangle$ 
are real.

Before introducing finite temperature corrections,
we summarize here some aspects of the tree level potential. 
First, it contains seven free parameters, but
imposing the stationarity conditions in $(H_{1}^{0},H_{2}^{0},N)= (v_{1},v_{2},x)$
with the constraint 
$ v_{1}^{2}+v_{2}^{2}\equiv v^{2} =(246 \; {\rm GeV})^{2}$ , 
we can express the soft masses in terms of the six parameters $\lambda$, $k$,
$A_{\lambda}$, $A_{k}$, $\tan\beta$ and $x$,
\bea  
m_{H_{1}}^{2} & = & \lambda(A_{\lambda}+kx)x \tan\beta
                -\lambda^{2}(x^{2} + v^{2} \sin^{2}\beta)
                -  \frac{g^{2}+g'^{2}}{4} v^{2} \cos 2\beta\nonumber\\
m_{H_{2}}^{2} & = & \lambda (A_{\lambda}+kx)\,x\, {\rm cotan} \beta
                -\lambda^{2} (x^{2} + v^{2} \cos^{2}\beta)
                +  \frac{g^{2}+g'^{2}}{4} v^{2} \cos 2\beta	\\
m_{N}^{2} & = & \lambda A_{\lambda}\frac{v^{2}}{2x}\sin 2\beta
                + k A_{k} x -\lambda^{2} v^{2} - 2 k^{2} x^{2}
                +  \lambda k v^{2} \sin 2\beta 	\nonumber\; .
\label{softmasses}
\eea
The above equations do not guarantee that $(v_{1},v_{2},x)$ is the global minimum of the tree level 
potential: for each choice of the parameters we must verify that this is indeed the case.

Concerning the physical masses of the 6 physical Higgs fields, one has 
to consider the matrix of the second derivatives of the tree-level
potential.
In order to avoid the formation of dangerous directions of 
instability in 
the space of charged and pseudoscalar Higgses, in the analysis of the strength
 of the transition for each choice of the parameters one has to verify the 
positiveness of their mass matrixes.

Finally, barring the possibility that the transition occurs along CP-violating 
directions, the potential becomes a function of three real scalar fields only,
$({\rm Re} H_1 ^0, {\rm Re} H_2 ^0, {\rm Re} N)$ which in the following 
will be denoted by $(\phi_1, \phi_2, \phi_3)$.

In our numerical analysis we  made use of 
the tree-level potential $V_F + V_D + V_{\rm soft}$, restricted to the three neutral scalar 
fields left, plus the one-loop corrections appearing at finite temperature.
Overall, we have six free parameters: the coupling parameters $\lambda$
and $k$; the soft-breaking mass terms  $A_{\lambda}$ and  $A_k$;
the zero-temperature vacuum expectation value  of  the singlet $x$ and $\tan\beta$.
The thermal corrections to the potential involve a linear term,
\be
\frac{1}{8}T^2 \sin (2\beta ) \: \lambda A_{\lambda} \phi_3	\; ,
\ee
and a bilinear one,
\be
\frac{T^2}{24} \left(  C_{H_1} \: \phi_1 ^2 +  C_{H_2} \: \phi_2 ^2 
 + C_{N} \: \phi_3 ^2 + 2 C_{12} \: \phi_1 \phi_2 \right)	\; ,
\ee
where the $C$'s are constants depending on $\lambda$, $k$ and $\tan \beta$ \cite{pietroni}.

\subsection{The landscape of minima}

Since we have now to deal with three scalars and six free parameters,
the NMSSM case is considerably more complicated
than the case of one light Higgs in the MSSM.
Besides increased computational difficulty, new features arise.
 
First of all, the strength of the transition is dominated by the 
tree-level trilinear term rather than by loop corrections, and
therefore it is not anymore 
directly related to the Higgses and stop masses  
(neither is related in a simple way to the six parameters of the 
tree level potential).

More in general, the landscape of local and global minima in this
large parameter space is more complicated.
Indeed, for some choices of the parameters 
the destabilization  does not even take place, while
for other values of the parameters, 
at $T = T_{\rm dest}$ the origin gets 
destabilized along a direction which does not connect the 
origin to the true vacuum, but to
a local minimum, separated from the global one by a barrier: 
in these cases the system first rolls down to
the new ``shifted'' false vacuum (second order transition),
and next it tunnels through the
barrier which separates it from the true vacuum (first order
transition). 
These two minima are
separated by a potential barrier in the three-dimensional field space
($\phi_1,\phi_2,\phi_3$); tunneling throughout the barrier will take place
through the trajectory in this field space which leads to the 
least Euclidean action (\ref{accion}) and this
in general will not be a straight line. 
Our approximation from now on is that this least-action trajectory
is in fact a straight line joining the two minima: in other words, 
after having exactly computed the
two minima, we reduce the problem to a one dimensional system along 
the direction joining these two points, 
thus simplifying the computation of bubble solutions. 
This perhaps leads to a small overestimate of
the strength of the transition.

In our analysis we have focussed on those regions of the parameter 
space which 
previous studies have shown to give rise to a large baryon asymmetry;
our strategy has been the following. First, we made a quicker scanning
of  wide regions in the six dimensional
parameters space. At this stage, it would be highly impractical to
reconstruct the bubble profile and compute $\alpha$ and $\beta$
for each value of the parameters
explored; instead, in order to have an idea of what are the most
interesting regions, we computed, at the degeneracy temperature
and (if it exists) at the destabilization temperature, 
the height of the potential barrier and the energy density 
difference between the false vacuum (possibly shifted, as discussed
above) and the true vacuum.

These first estimates allowed us to select the regions 
in the parameters space which
present a strong phase transition; in particular, it turns out that
 the most important parameters in the game
are the soft breaking mass terms $A_\lambda$ and $A_k$.
We then moved to a more accurate analysis restricted to some of the 
interesting regions found.

\begin{figure}
\centering\leavevmode\epsfxsize=5.3in\epsfbox{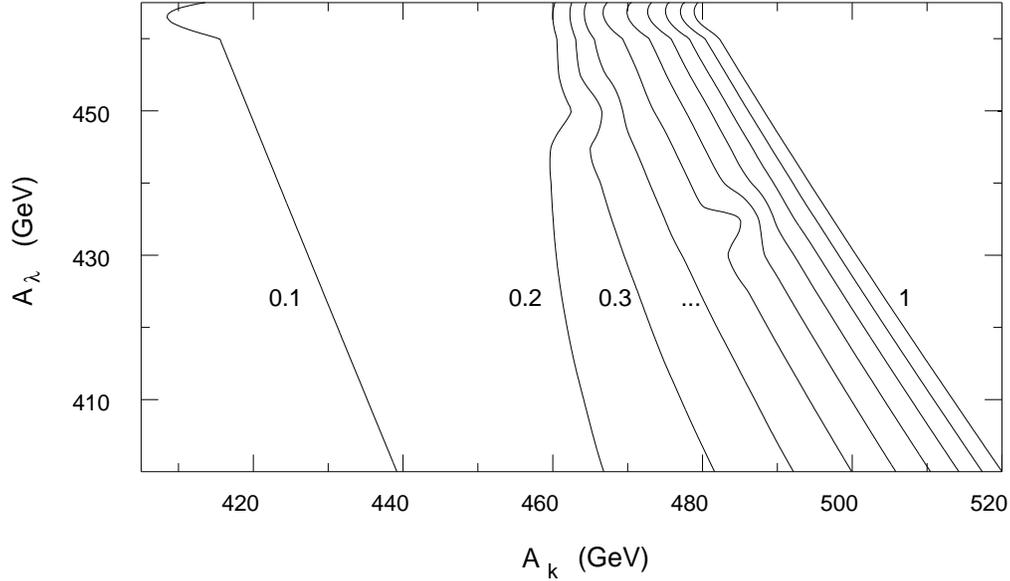}
\caption{Rough contour plot for $\alpha$ as a function of $ A_\lambda$ and $A_k$, for $x=350$ GeV, 
$\lambda=0.83$, $k=0.67$ and $\tan \beta=2$
. \label{NMSSMcontour}}
\end{figure}

\begin{figure}[H]
\centering\leavevmode\epsfxsize=4.5in\epsfbox{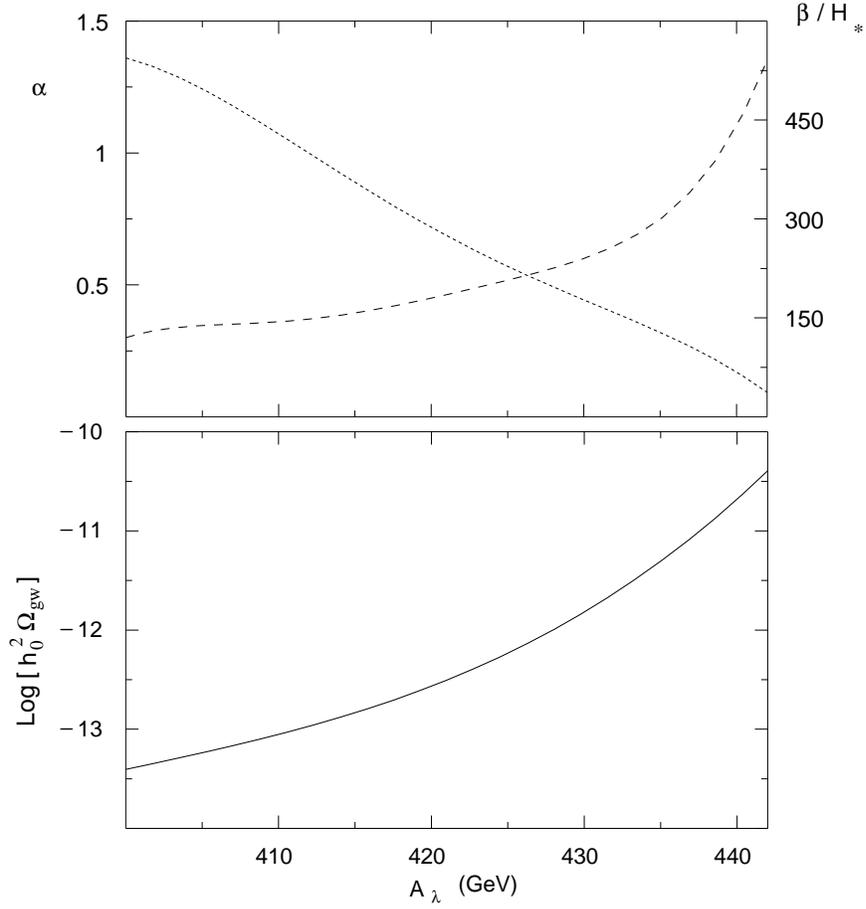}
\caption{Plot for \al\ (dashed line), $\beta / H_*$ (dotted line), 
and $\hogw$ (solid line)  from bubble collisions, as functions of
$A_\lambda$ for $A_k=480$ GeV, $x=350$ GeV, $\lambda=0.83$, $k=0.67$ and $\tan \beta=2$.
\label{NMSSMak}}
\end{figure}

\begin{figure}[H]
\centering\leavevmode\epsfxsize=4.5in\epsfbox{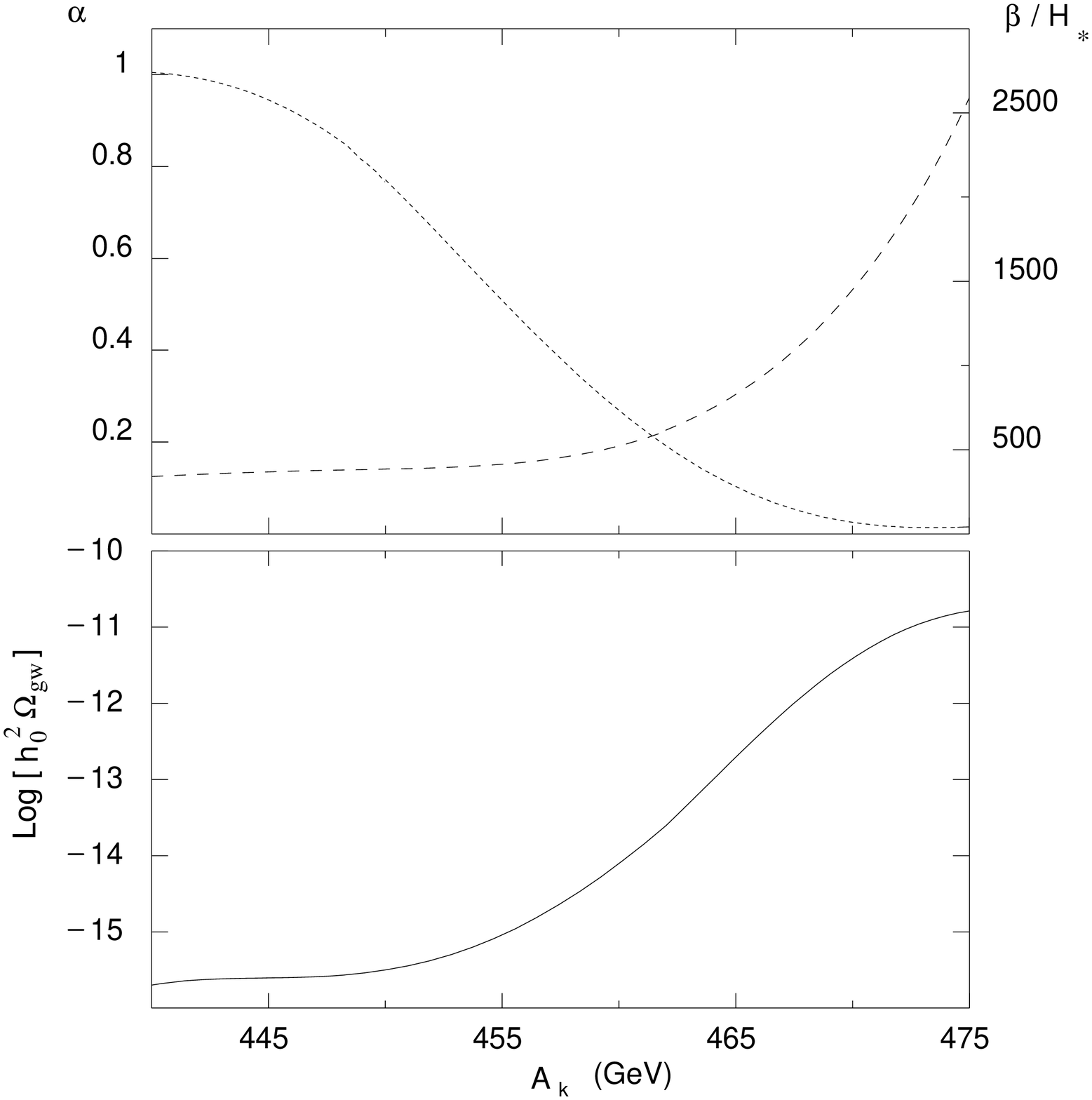}
\caption{Plot for \al\ (dashed line), $\beta / H_*$ (dotted line),
 and $\hogw$ (solid line)  from bubble collisions, as functions of
$A_k$ for $A_\lambda=450$ GeV, $x=350$ GeV, $\lambda=0.83$, $k=0.67$ and $\tan \beta=2$.
\label{NMSSMal}}
\end{figure}

The situation is now much more complicated than in the MSSM case, 
becuase of the higher dimensionality
of the Higgs space. Furthermore,
now the barrier is much higher than in the MSSM, 
and therefore the system has 
to experience an higher degree of supercooling before
the transition can take place. Thus, while in the MSSM the separations 
between the 
temperature of degeneracy for the two vacua $T_{\rm deg}$, the 
actual transition temperature $T_*$, and 
the origin-destabilization temperature $T_{\rm dest}$, 
are of order of a few GeVs,
in the NMSSM these intervals can be sensibly larger
(and indeed, as discussed above, for some ranges of the 
parameters the destabilization  can even be absent). 
This fact makes it harder the compute precisely the value
transition temperature $T_*$, and thus of $\alpha$.
In order to get a contour plot for the function $\alpha (A_\lambda,A_k
)$ we proceeded as follows. The transition
takes place when $S_3(T)/T \simeq 140$, $S_3$ being the sum of a kinetic term and a potential one 
(see (\ref{accion})). These two terms are likely to be of the same order if evaluated on the bubble-like
solution of the equations of motion, and thus $S_3 (T) \propto \int \! dr \, r^2 \, V(\phi_b,T)$. 
The integral of the potential extended to the bubble
 receives a contribution  mainly from the wall
proportional to the maximum $V_{\rm max}$ of the potential and a contribution from the inside of
the bubble proportional to the minimum $V_{\rm min}$, {\em i.e.} $S_3 (T) \sim A \cdot V_{\rm max}(T) + 
B \cdot V_{\rm min} (T)$, with $A$ and $B$ depending on the exact solution $\phi_b (r)$. 
The condition to
determine $T_*$ thus becomes $A \cdot V_{\rm max}(T_*) + B \cdot V_{\rm min} (T_*) \sim 140 \, T_*$.
We assumed constant $A$ and $B$ in each small region we explored, thus fitting
their value with several exact solutions $\phi_b (r)$.
As in the MSSM case, once this (approximate) transition temperature $T_*$ is evaluated, one 
easily computes $\alpha$ from its definition (\ref{defalpha}). 
A result of this procedure is summarized in 
fig.~\ref{NMSSMcontour}, which shows a rapid growth of $\alpha$ versus $A_\lambda$ and $A_k$, with 
$x=350$ GeV, $\lambda = 0.83$, $k=0.67$ and $\tan \beta = 2$.
Next we moved to computing exactly $T_*$, $\alpha$ and $\beta / H_*$ for specific values
of $A_\lambda$ and $A_k$ in the region of fig.~\ref{NMSSMcontour}, again by numerically computing
the temperature for which the bubble solution $\phi_b(r)$ gives $S_3 (T) / T \simeq 140$
and then applying relations (\ref{defalpha}) and (\ref{defbsuh}). 
Typical  results are shown in Figs.~\ref{NMSSMak} and \ref{NMSSMal}, 
where we plot \al, $\beta / H_*$ and $\hogw$ as functions of $A_\lambda$ and $A_k$.

\begin{figure}
\centering\leavevmode\epsfxsize=4.4in\epsfbox{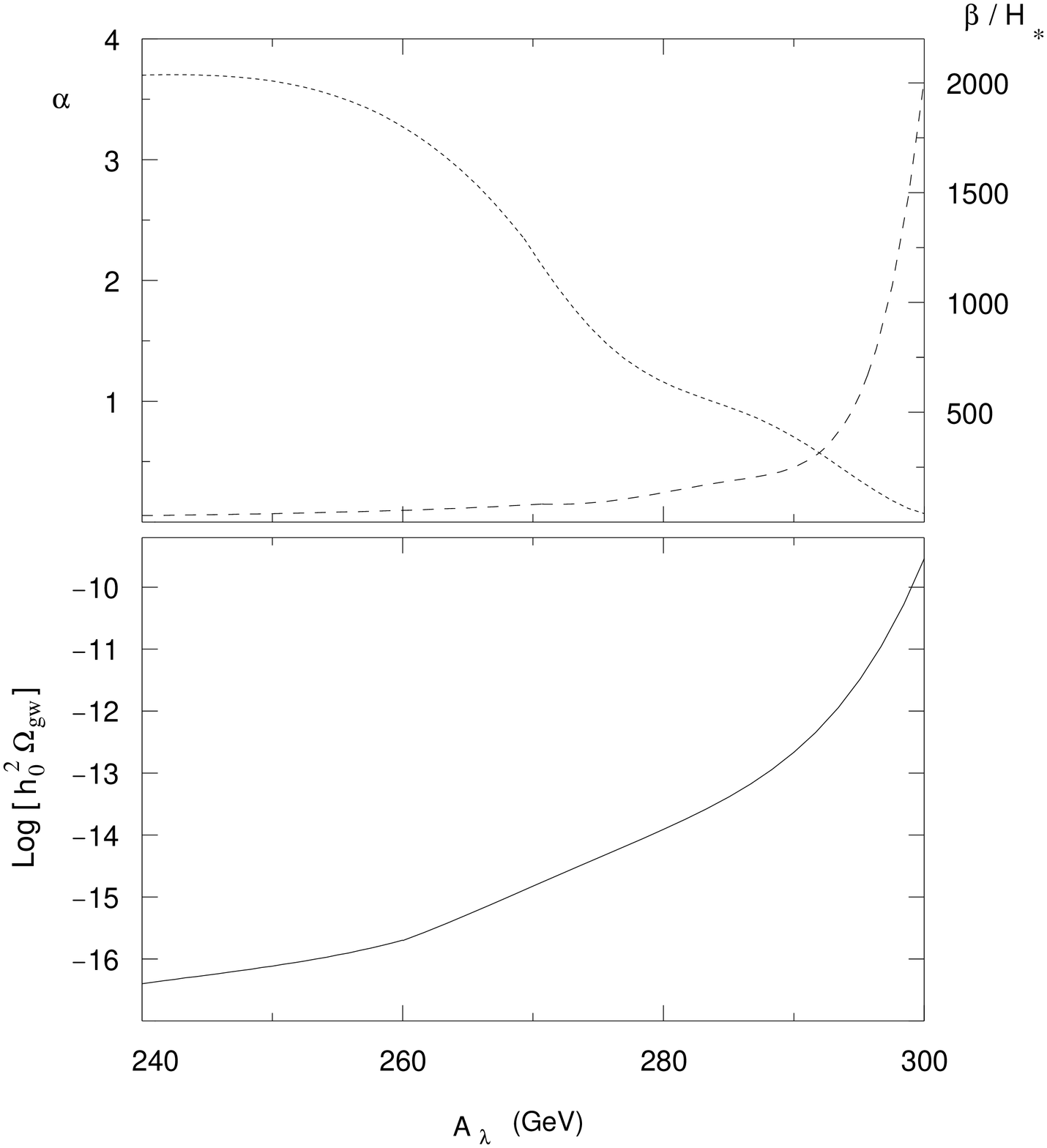}
\caption{Plot for \al\ (dashed line), $\beta / H_*$ (dotted line),
 and $\hogw$  from bubble collisions (solid line), as functions of
$A_\lambda$ for $A_k=167$ GeV, $x=300$ GeV, $\lambda=0.523$, $k=0.37$ and $\tan \beta=2$.
\label{NMSSMak167}}
\end{figure}

Note that, since the
transition in the NMSSM is strongly first order,  it is correct to
use the expression for $v(\alpha )$ computed for detonation 
bubbles~\cite{kamionkowski}.
The results shown in the figures are not peculiar of the values of the parameters chosen: similar results
can be found in completely different regions of the parameters space, 
see for instance fig.~\ref{NMSSMak167}.  

\begin{figure}
\centering\leavevmode\epsfxsize=4.4in\epsfbox{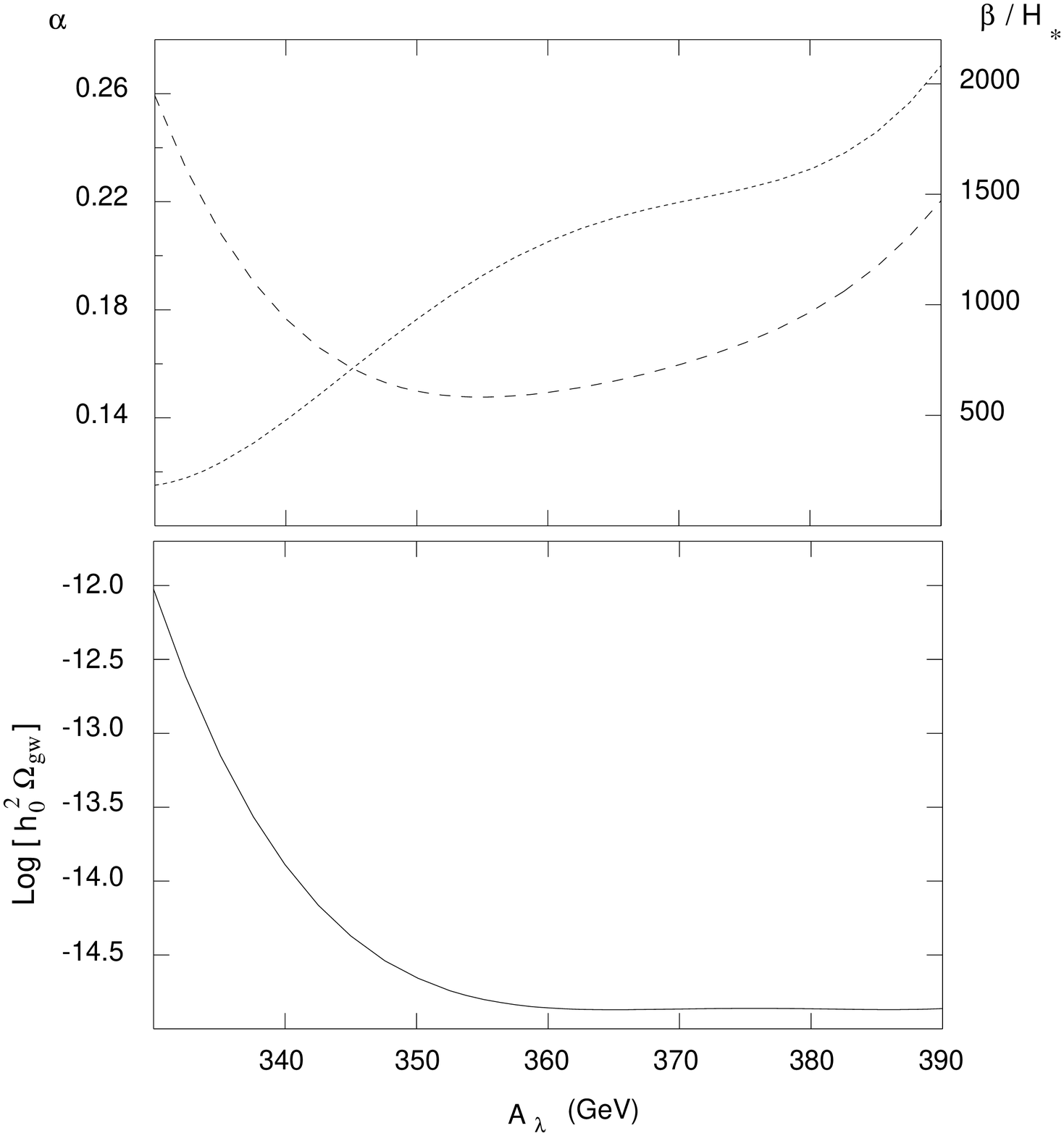}
\caption{Plot for \al\ (dashed line), $\beta / H_*$ (dotted line),
 and $\hogw$  from bubble collisions, (solid line) as functions of
$A_k$ for $A_\lambda=450$ GeV, $x=350$ GeV, $\lambda=0.83$, $k=0.67$ and $\tan \beta=2$.
\label{NMSSMtipo2}}
\end{figure}

\begin{figure}
\centering\leavevmode\epsfxsize=5.2in\epsfbox{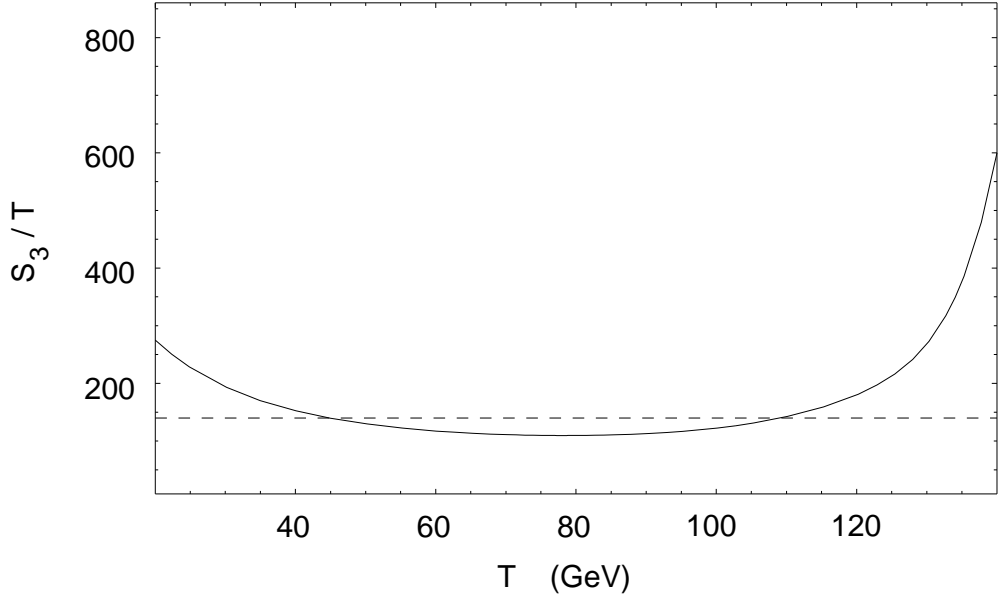}
\caption{Behaviour of the action $S_3/T$ versus $T$ in an ``extremal'' region of the parameters-space.
The horizontal dashed line indicates the value $S_3/T \simeq 140$ that must be reached in order that the transition 
takes place.
\label{s3sutnmssm}}
\end{figure}

An example of a region of the parameter space in which the transition happens through the nucleation of
bubbles subsequent to a smooth roll-down is shown in fig.~\ref{NMSSMtipo2}.
\begin{figure}[H]
\hspace*{-12mm}
\centering\leavevmode\epsfxsize=4in\epsfbox{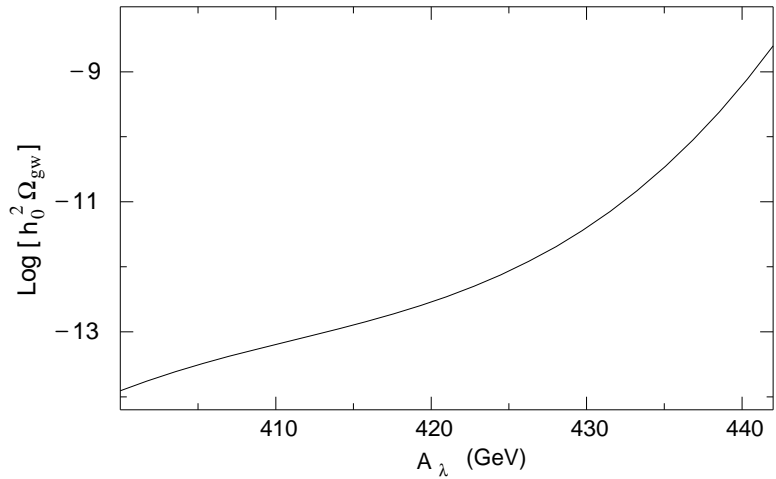}
\caption{$\hogw$ from turbolence
as a function of $A_\lambda$ for $A_k=480$ GeV, $x=350$ GeV, $\lambda=0.83$, $k=0.67$ and $\tan \beta=2$.
\label{turboN1}}
\end{figure}
\begin{figure}[H]
\hspace*{-11mm}
\centering\leavevmode\epsfxsize=4.15in\epsfbox{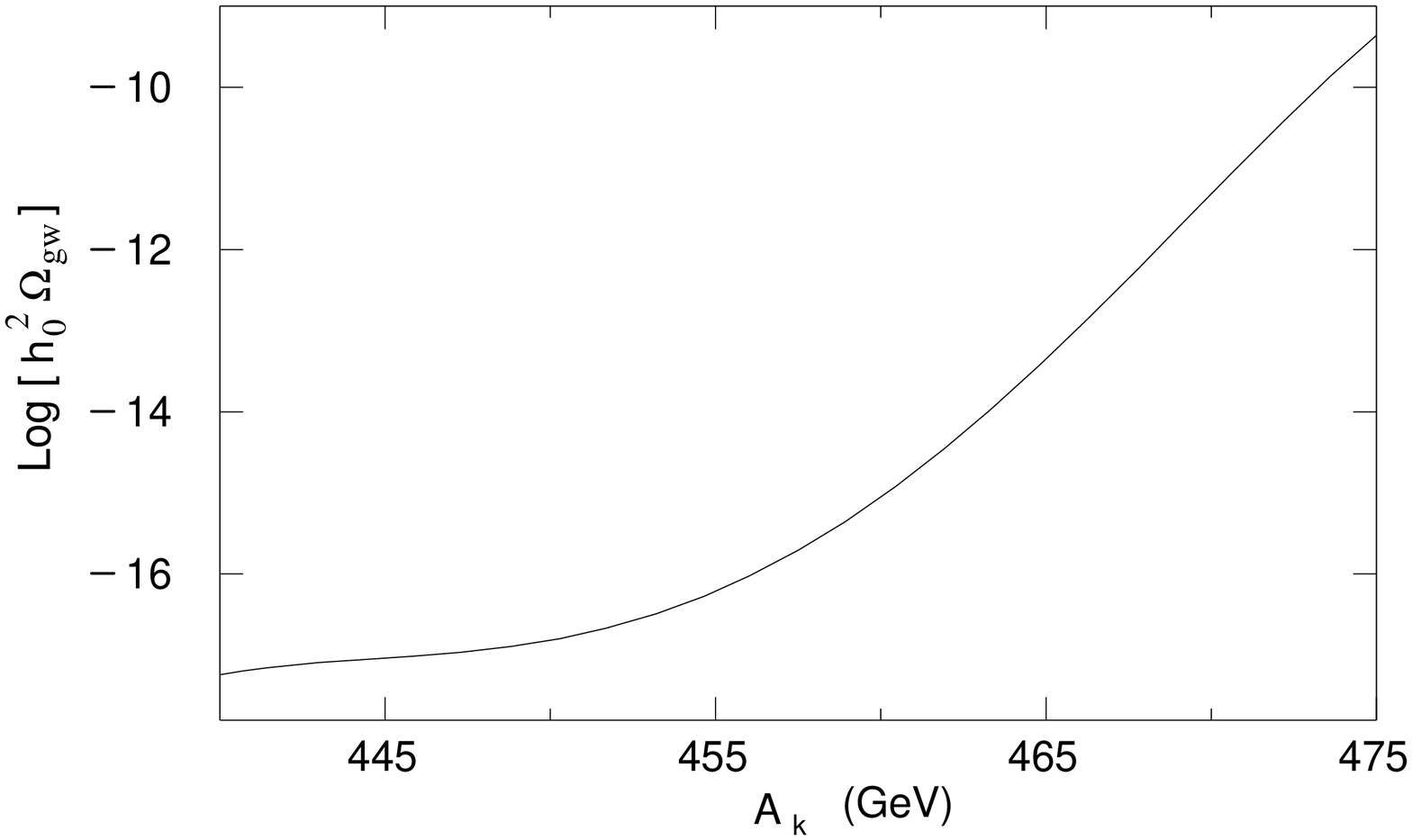}
\caption{$\hogw$ from turbolence as a function of $A_k$ for $A_\lambda=450$ GeV, $x=350$ GeV, $\lambda=0.83$, $k=0.67$ and $\tan \beta=2$.
\label{turboN2}}
\end{figure}
\begin{figure}
\hspace*{-11mm}
\centering\leavevmode\epsfxsize=4in\epsfbox{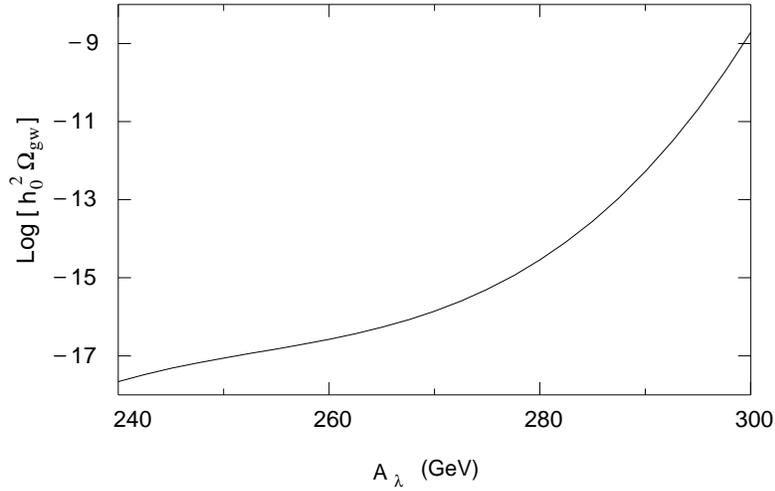}
\caption{$\hogw$ from turbolence as a function of $A_\lambda$ for $A_k=167$ GeV, $x=300$ GeV, $\lambda=0.523$, $k=0.37$ and $\tan \beta=2$.
\label{turboN3}}
\end{figure}
\begin{figure}
\hspace*{-10mm}
\centering\leavevmode\epsfxsize=4.1in\epsfbox{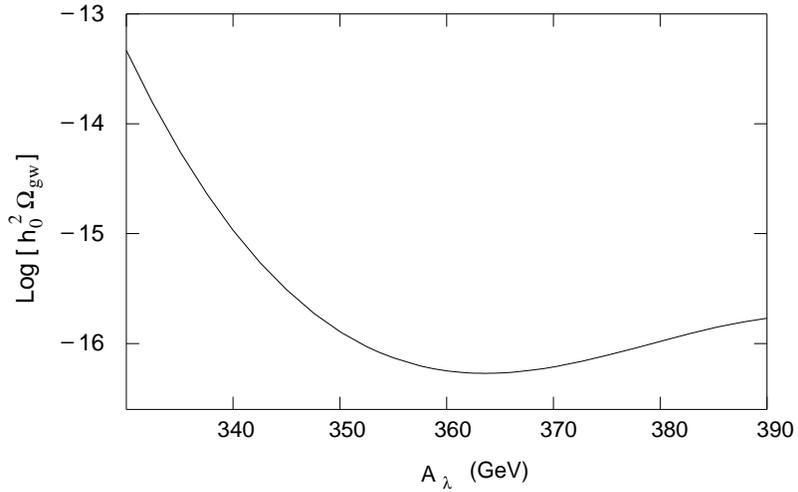}
\caption{$\hogw$ from turbolence as a function of $A_k$ for $A_\lambda=450$ GeV, $x=350$ GeV, $\lambda=0.83$, $k=0.67$ and $\tan \beta=2$.
\label{turboN4}}
\end{figure}
In figs.~\ref{NMSSMak167} and \ref{NMSSMal}
the intensity of the produced gravity waves  grows rapidly as a function
of the soft breaking mass terms. 
However, $A_\lambda$ and $A_k$ cannot be extended beyond the
values shown in the figures: beyond such values 
the condition $S_3(T_*)/ T_* \simeq 140$ is never satisfied and the
transition does not take place.
This is due to the presence of a potential barrier also at zero temperature, generated by
trilinear terms in the tree-level potential: as discussed in section \ref{2c}, the Euclidean action $S_3(T)$ computed
on the bubble solution decreases from $+ \infty$ to 0 in going from $T=T_{\rm deg}$ to $T=T_{\rm dest}$; if
the origin never gets
destabilized -- as in the cases we are considering here -- one
has $S_3(T) \to {\rm const} \neq 0$
as $T \to 0$. The value of this constant is related to the heigth of the zero-temperature potential barrier,
and it is possible that the quantity $S_3(T)/ T$ never 
reaches the critical value $\simeq 140$. This behavior is
illustrated in
fig.~\ref{s3sutnmssm}~~\footnote{Notice that in these ``extremal'' cases there exists a large range of $T$ in 
which $S_3(T)/ T$ is low
and slow-varying: for such cases the integral (\ref{integral}) must be computed
numerically and leads to a little less restrictive condition for the temperature of the transition, however always
of the form $S_3(T_*)/ T_* \simeq 140 - 145$.}.
It is clear that moving to these extreme regions 
makes $\beta / H_*$ close to zero, because it is proportional
to the slope of $S_3 / T$ computed at the ``crossing'' $S_3 / T \simeq 140$ (see fig.~\ref{s3sutnmssm}). By
physical reasons, one cannot accept values of $\beta / H_*$ smaller than 1: for such values the transition would be
so slow that it doesn't complete in a Hubble time, and clearly the computations that lead to (\ref{omega}) are
not valid in such a case. 
Our best results give therefore values of $\hogw\sim 10^{-10}$, peaked
at a frequency, obtained from eq.~\ref{peakf}, of approximately 10~mHz.

\subsection{Gravitational waves from turbolence in the NMSSM}
We now estimate the amount of GWs produced by turbolence in the case of the NMSSM for the specific regions
of the parameter space studied in the previous section.
Substituting in eq. (\ref{turbo}) the values of $\alpha$ and $\beta / H_*$ 
shown in figs.~\ref{NMSSMak}, \ref{NMSSMal}, \ref{NMSSMak167} and \ref{NMSSMtipo2} 
we obtain for $\hogw$ the plots of figs.~\ref{turboN1}, \ref{turboN2}, \ref{turboN3} and \ref{turboN4}.

We recall again that, due to the difficulties in estimating accurately
the production of GWs from turbolence, these plots should be taken
only as indicative. Still, they suggest that turbolence can be a
really powerful source of GWs, indeed more powerful than the
better studied mechanism of bubble collisions. In special points of the
parameter space, we find values of $\hogw$ even of order $10^{-9}$,
which would correspond to a signal-to-noise ratio of order 100 for LISA!
It is clear that the production of GWs by turbolence deserves further
studies.

\section{Conclusions \label{sez6}}

The production of GWs at the electroweak
scale is strongly dependent on the model used and on the parameters of
the model. This is an expected consequence of the well known fact that
the strength of the phase transition is itself strongly dependent. 
Our aim was to compare the production of GWs with the reference value
given by the sensitivity of LISA. In the Standard Model, for the
experimentally allowed values of the Higgs mass, there is no
electroweak phase transition at all, and therefore no production of
GWs. We have therefore turned our attention to supersymmetric
extensions. For the MSSM the answer is again negative. The results
never exceed values of order $\hogw\sim 10^{-16}$, five orders of magnitudes
below the sensitivity of LISA.

More encouraging results can be obtained in the Next-to-Minimal
Supersymmetric Standard Model, basically because in this case the
transition is due to tree level couplings, rather than being
radiatively induced. In this case, while in many regions of the
parameter space the signal is still neglegible, there are also 
region where one can get 
 $\hogw\sim  10^{-10}$ from bubble nucleation and
even $\hogw\sim  10^{-9}$ from turbolence, 
with a peak frequency around 10~mHz.
A background of this intensity would be within the
reach of LISA. 

While regions of parameter space with such large values are, generically
speaking, quite small, it must be observed that 
the condition for an intense
GW production is  the same as the condition for the generation
of the baryon asymmetry at the electroweak scale.

\end{document}